\documentclass[reprint,superscriptaddress,amsmath,amssymb,aps,prb,floatfix]{revtex4-2}

\usepackage{siunitx}
\usepackage{makecell}
\usepackage[colorlinks,linkcolor=black,anchorcolor=blue,urlcolor=blue,citecolor=blue,unicode]{hyperref}
\usepackage{multirow}
\usepackage[normalem]{ulem}
\usepackage{booktabs,array}
\usepackage{listings}
\usepackage{import}
\usepackage{xcolor}
\usepackage{framed}
\usepackage{mdwlist}
\usepackage{diagbox}
\usepackage{bm}
\usepackage{graphicx}
\usepackage{dcolumn}
\usepackage{algorithm}
\usepackage{algpseudocode}
\usepackage{csquotes}
\usepackage{ifthen}
\usepackage{threeparttable}
\usepackage{float}
\usepackage{tabularx}
\usepackage{CJK}
\usepackage{indentfirst}
\usepackage[caption=false]{subfig}

\begin{document}

\title{Finite-temperature adiabatic-connection reduced density-matrix functional theory: a unified framework for i-DMFT and TAO-DFT
}

\author{Jingang Han}
\affiliation{Institute of Physics, Chinese Academy of Sciences, Beijing 100190, China}
\affiliation{School of Physical Sciences, University of Chinese Academy of Sciences, Beijing 100049, China}

\author{Huanjing Gong}
\affiliation{Institute of Physics, Chinese Academy of Sciences, Beijing 100190, China}

\affiliation{School of Physical Sciences, University of Chinese Academy of Sciences, Beijing 100049, China}
\author{Kai Luo}
\email{kluo@njust.edu.cn}
\affiliation{Department of Applied Physics, Nanjing University of Science and Technology, Nanjing 210094, China}

\author{Hongming Weng}
\email{hmweng@iphy.ac.cn}
\affiliation{Institute of Physics, Chinese Academy of Sciences, Beijing 100190, China}
\author{Xinguo Ren}
\email{renxg@iphy.ac.cn}
\affiliation{Institute of Physics, Chinese Academy of Sciences, Beijing 100190, China}

\newcommand{\XR}[1]{\color{red}{\bf #1 }}
\newcommand{\XB}[1]{\color{blue}{ #1 }}
\newcommand{\XY}[1]{\color{yellow}{\bf #1 }}
\newcommand{\XG}[1]{\color{green}{\bf #1 }}

\definecolor{clorange}{RGB}{200,90,0}
\newcommand{\CL}[1]{{\color{clorange}\sffamily\small$\blacktriangleright$[CL: #1]$\blacktriangleleft$}}
\definecolor{clgreen}{RGB}{0,130,60}
\newcommand{\GX}[1]{{\color{clgreen}#1}}
\newcommand{\CLnote}[1]{{\color{clorange}\sffamily\small$\blacktriangleright$[CL-r2: #1]$\blacktriangleleft$}}

\begin{abstract}
An accurate description of chemical bond breaking requires capturing strong/static electron correlation, which is highly challenging for single-reference self-consistent-field methods such as Hartree--Fock and Kohn--Sham density-functional theory (DFT). Introducing fractional occupations substantially improves this situation, as exemplified by the recently developed information-density-matrix-functional theory (i-DMFT) and thermally-assisted-occupation DFT (TAO-DFT). Although practically useful, these two approaches are formulated within rather different theoretical frameworks and are somewhat semi-empirical in nature, and their relationship has remained largely obscure. Here we introduce a finite-temperature adiabatic-connection reduced density-matrix functional theory (FT-AC-RDMFT), within which i-DMFT and TAO-DFT emerge as different reductions of a unified framework. The effective eigenvalue equation for the natural orbitals and the Fermi--Dirac form of the occupation numbers follow directly from the variational principle, and a thermal adiabatic connection relates the empirical entropy coefficient of i-DMFT to a characteristic temperature of the auxiliary non-interacting system. Within this framework, i-DMFT corresponds to the exact-exchange-based reduction, whereas TAO-DFT is obtained when the exact-exchange term is replaced by a density-based exchange-correlation functional. Numerical calculations confirm that i-DMFT, apart from a constant, coincides with the exact-exchange realization of the framework, and show that the improvement in dissociation behavior brought by finite-temperature occupations is not restricted to exact exchange but also arises with hybrid and semilocal density functionals. For the strongly correlated oxide V$_2$O$_3$, the finite-temperature occupations substantially reduce the severely overestimated non-magnetic LDA bulk modulus and bring it close to the experimental value. These results establish FT-AC-RDMFT as a common theoretical foundation for finite-temperature occupation-based electronic-structure methods in both molecular and solid-state systems.
\end{abstract}

\maketitle

\section{Introduction}

A central challenge in electronic-structure theory is to achieve an accurate description of electron correlation without sacrificing computational tractability for realistic systems. Wave-function-based many-body methods, such as configuration interaction (CI) and coupled-cluster (CC) theory, can achieve high accuracy, but their computational cost grows rapidly with system size~\cite{sherrill1999configuration,bartlett2007coupled}. By contrast, Kohn--Sham density functional theory (DFT)~\cite{PhysRev.136.B864,PhysRev.140.A1133,parr1995density,xu2005extended,cohen2008insights,dreizler2012density,su2018describing} has become the standard workhorse of quantum chemistry and condensed-matter physics because of its excellent balance between accuracy and computational efficiency. However, conventional density-functional approximations often struggle in situations with strong/static correlation, such as bond dissociation and near-degeneracy situations~\cite{cohen2008insights,cohen2008fractional}.

Reduced density-matrix functional theory (RDMFT) provides an appealing alternative framework. Since Gilbert's extension of the Hohenberg--Kohn theorem to the one-body reduced density matrix (1-RDM), together with related early foundational works, RDMFT has rested on a formal foundation in which the ground-state energy is treated as a functional of the 1-RDM rather than only the density~\cite{gilbert1975hohenberg,donnelly1979fundamental,donnelly1978elementary,nguyen1985variation,berrondo1975energy,levy1979universal,zumbach1985density,muller1984explicit,goedecker1998natural}. In recent years, it has attracted renewed interest for its potential to describe strongly correlated systems~\cite{pernal2005effective,gritsenko2005improved,rohr2008density,sharma2008reduced,lathiotakis2009density,piris2010communications,sharma2013spectral,schade2017reduced,schilling2018communication,schilling2019diverging,cioslowski2020one,piris2021global,yao2021handling,ai2023short,gibney2022comparison,gibney2023universal,cartier2024exploiting}. The 1-RDM admits the spectral representation
\begin{equation}
\gamma(x,x')=\sum_i n_i\, \varphi_i(x)\varphi_i^*(x'),
\end{equation}
where $\{\varphi_i\}$ are the natural orbitals (NOs) and $\{n_i\}$ are the natural occupation numbers (ONs), satisfying the ensemble-$N$-representability conditions $0 \le n_i \le 1$ and $\sum_i n_i = N$. This structure makes RDMFT a natural framework for describing static correlation, where an explicit treatment of fractional occupations is essential.

Despite this conceptual appeal, the development of RDMFT has been much slower than originally hoped. An important reason is the lack of sufficiently accurate and broadly transferable approximate functionals. Another equally serious obstacle is the difficulty of practical minimization. In conventional zero-temperature RDMFT optimizations, effective-Hamiltonian-based approaches are hindered because not all matrix elements required in the optimization are well defined~\cite{pernal2005effective,piris2009iterative}, while in direct-minimization approaches the optimization of natural orbitals and occupation numbers tends to compete. Considerable effort has recently been devoted to improving these minimization schemes, and their efficiency has been substantially advanced~\cite{yao2021handling,yao2022explicit,yao2024enhancing,lew2025advancing,luo2026iterative}. Nevertheless, the optimization remains more involved and less robust than a standard Kohn--Sham DFT self-consistent-field calculation.
As a result, even though RDMFT is formally attractive, its practical use has long been limited by both functional limitations and the numerical difficulty of minimizing the RDMFT functional efficiently.

A key conceptual difficulty behind this situation is that the exact ground-state 1-RDM of an interacting system is generally nonidempotent and therefore cannot, in general, be represented by a non-interacting ground state. This makes it difficult to construct a simple self-consistent one-particle equation of the Kohn--Sham type directly within zero-temperature RDMFT~\cite{BALDSIEFEN2013114}. Finite-temperature RDMFT (FT-RDMFT) partially addresses this by introducing an auxiliary non-interacting system in grand-canonical equilibrium, which can in principle represent an interacting 1-RDM arbitrarily closely~\cite{BALDSIEFEN2013114,baldsiefen2015reduced}; in practice, however, its solution still requires a self-consistently determined pseudo-Hamiltonian and remains cumbersome to apply.

A different and practically very successful route to improve the description of static correlation is to introduce fractional occupations directly at the mean-field level. Information-density-matrix-functional theory (i-DMFT) introduces an entropy-related contribution and has been shown to yield much improved dissociation curves.
In i-DMFT, the central working assumption is that the correlation energy associated with the two-particle cumulant can be represented, to leading order, by an information-entropy functional of the natural occupation numbers~\cite{wang2021information,wang2022scf,irimia2023self,liu2024potential}. This assumption leads to an energy functional of Hartree--Fock form supplemented by an entropic term, which favors fractional occupations when near-degenerate orbitals appear. As a result, i-DMFT has been shown to describe molecular bond dissociation with a quality that is difficult to obtain from conventional single-reference methods, while retaining a computational structure similar to a standard self-consistent-field calculation. In particular, the natural orbitals are obtained by diagonalizing a Fock-like effective Hamiltonian and the occupation numbers follow a Fermi--Dirac form, so that the severely coupled optimization of natural orbitals and occupation numbers in conventional RDMFT is avoided.

Thermally assisted occupation density functional theory (TAO-DFT) addresses static correlation from a different premise~\cite{chai2012density}. Formulated as a density-based theory, it assumes that the zero-temperature density of an interacting system can be represented by the thermal-equilibrium density of an auxiliary non-interacting Kohn--Sham system at a fictitious temperature. The role of this temperature is therefore to generate fractional Kohn--Sham occupations within a density-functional partition of the energy, together with density-based exchange-correlation approximations and a finite-temperature non-interacting free-energy correction. The exchange treatment in TAO-DFT ranges from local-density (LDA) and semilocal (GGA) approximations~\cite{chai2012density,chai2014thermally} to hybrid schemes incorporating exact exchange, including global and range-separated variants~\cite{chai2017role}. This is conceptually distinct from i-DMFT, which starts from an orbital- and occupation-dependent 1-RDM energy expression with exact exchange and introduces the entropy term as an information-theoretic approximation to the correlation energy associated with the two-particle cumulant. Thus, although both approaches use a temperature-inspired occupation structure to describe static correlation, their basic variables, functional decompositions, and underlying assumptions are different. In particular, the entropy contribution in TAO-DFT is absorbed into the non-interacting kinetic free energy, whereas in i-DMFT the entropy term is used to model the interaction-related cumulant correlation energy.

A common feature of FT-RDMFT, i-DMFT, and TAO-DFT is that an entropy term of the occupation numbers appears in each of them. Its role, however, differs markedly among the three: in FT-RDMFT it is the thermodynamic entropy of the grand-canonical ensemble, in i-DMFT it is used to model the correlation energy, and in TAO-DFT it is absorbed into the non-interacting kinetic free energy. This raises the question of whether these seemingly different entropy contributions can be understood on a common footing, and more broadly whether i-DMFT and TAO-DFT should be regarded as fundamentally different constructions or as different reductions of a single variational framework. Clarifying their mutual relationship, and their relationship to RDMFT, requires analyzing the origin of the Fermi--Dirac occupations and the distinction between exact-exchange-based and density-based formulations from the perspective of the 1-RDM.

In this work, we address these issues by starting from FT-RDMFT and reformulating the theory in a way that naturally leads to an FT-AC-RDMFT (adiabatic-connection FT-RDMFT) framework, where the adiabatic connection links the finite-temperature non-interacting reference to the zero-temperature interacting energy (Sec.~\ref{sec:adiabatic}). By varying the energy functional with respect to the natural orbitals and occupation numbers, one obtains both the effective eigenvalue equation and the Fermi--Dirac form of the occupations directly from the 1-RDM variational principle. This shows explicitly that the finite-temperature occupation structure is not imposed by hand, but emerges naturally within the theory.

This framework directly answers the question raised above: i-DMFT and TAO-DFT are not fundamentally different constructions, but two reductions of a single variational theory, distinguished by whether the auxiliary non-interacting system reproduces the full 1-RDM or only the density. This is reflected, at the level of the working equations, in whether the effective single-particle potential is nonlocal or local, and correspondingly in the natural choice of exact versus density-based exchange. The entropy terms that play such different roles in the two methods thus share a common origin in the introduction of a finite-temperature non-interacting reference, even though this reference is constrained differently in the two cases. The resulting working equations share the structure of the finite-temperature Hartree--Fock or Kohn--Sham equations and can be solved by the same techniques, a favorable computational feature already exploited in i-DMFT~\cite{wang2022scf}.

We then present self-consistent calculations for representative molecules and solids. We first confirm numerically that i-DMFT is equivalent to finite-temperature Hartree--Fock (FT-HF), up to a constant energy shift, when the same temperature parameter is used, and that the empirical entropy coefficient of i-DMFT can be identified with the temperature of the FT-HF free-energy functional through the adiabatic-connection analysis, FT-HF being the first-order, exact-exchange realization of FT-RDMFT~\cite{baldsiefen2015reduced}.
We further show that the favorable dissociation behavior associated with finite-temperature occupations is not restricted to exact exchange, but also appears with hybrid and semilocal density functionals, and extends from molecules to solids. For the strongly correlated oxide V$_2$O$_3$, in particular, the finite-temperature occupations substantially reduce the severely overestimated non-magnetic LDA bulk modulus and bring it close to the experimental value. These results indicate that the beneficial role of the finite-temperature occupation structure is general, across both exchange treatments and molecular and solid-state systems.

The rest of the paper is organized as follows. Section~\ref{sec:theory} presents the theoretical framework: the finite-temperature RDMFT setting, the thermal adiabatic connection that relates it to i-DMFT, the energy decomposition that connects it to TAO-DFT, and the resulting variational equations for the natural orbitals and occupation numbers, together with the approximations used in practical calculations. Section~\ref{sec:results} reports self-consistent calculations for molecules and solids, including the equivalence between i-DMFT and FT-HF, the comparison across exact-exchange, hybrid, and semilocal functionals, and the applications to solid hydrogen and V$_2$O$_3$. Section~\ref{sec:discussion} discusses the implications of these results, and Section~\ref{sec:summary} summarizes the work.

\section{Theory}
\label{sec:theory}

\subsection{Finite-temperature RDMFT}
\label{sec:ftrdmft}

RDMFT provides a variational framework in which the ground-state properties of an interacting many-electron system are described in terms of the one-body reduced density matrix (1-RDM), $\gamma(x,x')$, where $x=(\mathbf{r},\sigma)$ denotes a combined space-spin coordinate. We consider a system with Hamiltonian $\hat{H}=\hat{T}+\hat{V}_{\mathrm{ext}}+\hat{W}$, where $\hat{T}$ is the kinetic energy, $\hat{V}_{\mathrm{ext}}$ is the external potential, and $\hat{W}$ is the two-body interaction. The total energy can be expressed as a functional of $\gamma$,
\begin{equation}
E[\gamma] = F[\gamma] + \int \gamma(\mathbf{r},\mathbf{r}')\, v_{\mathrm{ext}}(\mathbf{r}',\mathbf{r}) \, d\mathbf{r}\, d\mathbf{r}',
\end{equation}
where $\gamma(\mathbf{r},\mathbf{r}')=\sum_{\sigma}\gamma(\mathbf{r}\sigma,\mathbf{r}'\sigma)$ is the spin-summed 1-RDM, which is the relevant quantity because the external potential considered here is spin-independent. The universal functional can be decomposed into its kinetic and interaction parts,
\begin{equation}
F[\gamma] = T[\gamma] + V_{ee}[\gamma].
\end{equation}
In the present work we restrict ourselves to local external potentials, i.e.,
\begin{equation}
v_{\mathrm{ext}}(\mathbf{r}',\mathbf{r}) = v_{\mathrm{ext}}(\mathbf{r})\,\delta(\mathbf{r}-\mathbf{r}'),
\end{equation}
so that the external-potential term reduces to $\int \rho(\mathbf{r})\, v_{\mathrm{ext}}(\mathbf{r})\, d\mathbf{r}$, with $\rho(\mathbf{r})=\gamma(\mathbf{r},\mathbf{r})$.

At zero temperature, it is well known that the ground-state 1-RDM of an interacting system cannot, in general, be represented by a non-interacting system, due to its non-idempotent nature. This limitation can be overcome within finite-temperature RDMFT (FT-RDMFT), where one considers a non-interacting system in grand canonical equilibrium. In this case, the occupation numbers follow the Fermi--Dirac distribution, ensuring that any ensemble-$N$-representable 1-RDM can be approximated arbitrarily closely~\cite{mermin1965thermal,baldsiefen2015reduced,BALDSIEFEN2013114}.

Within this framework, one introduces the grand potential functional of a non-interacting reference system,
\begin{equation}
\Omega_0[\gamma] = T_0[\gamma] + V_{\mathrm{aux}}[\gamma] - \mu N[\gamma] - \frac{1}{\beta} S_0[\gamma],
\label{eq:omega0}
\end{equation}
Here $V_{\mathrm{aux}}[\gamma]$ denotes the one-body potential energy of the auxiliary finite-temperature reference system, rather than the physical external-potential contribution in Eq.~(2). In the present 1-RDM formulation, the corresponding auxiliary one-body potential should be understood as an effective potential acting on the full 1-RDM and may therefore be nonlocal, even when the physical external potential of the original interacting system is local.
In the grand-potential expression above, $\mu$ is the chemical potential, $\beta = 1/\theta$ is the inverse temperature, and $S_0[\gamma]$ is the non-interacting entropy functional given by
\begin{equation}
S_0[\{n_i\}] = - \sum_i \left[ n_i \ln n_i + (1-n_i)\ln(1-n_i) \right].
\label{eq:S0}
\end{equation}
The kinetic energy of the non-interacting system reads
\begin{equation}
T_0[\gamma] = -\frac{1}{2} \sum_i n_i \int \varphi_i^*(\mathbf{r}) \nabla^2 \varphi_i(\mathbf{r}) \, d\mathbf{r}.
\label{eq:T0}
\end{equation}
Because the kinetic energy is a one-body operator, the kinetic energy of any state with 1-RDM $\gamma$ is given exactly by $T_0[\gamma]$, i.e.\ $T[\gamma]=T_0[\gamma]$; in RDMFT there is thus no kinetic correlation energy.

This finite-temperature formulation makes it possible to represent an interacting 1-RDM through an auxiliary non-interacting reference. To connect this reference to the zero-temperature interacting energy $E[\gamma]$, we introduce in the following subsection an adiabatic connection, which provides the starting point for the FT-AC-RDMFT working equations obtained below.

\subsection{Thermal adiabatic connection and reduction to i-DMFT}\label{sec:adiabatic}

Once the non-interacting auxiliary system is established, an explicit connection between it and the zero-temperature interacting system can be obtained by introducing a coupling-constant path at fixed 1-RDM, along which the temperature is varied simultaneously. Whereas previously formulated finite-temperature adiabatic connections are constructed at fixed temperature~\cite{baldsiefen2015reduced}, the path used here varies $\theta(\lambda)$ from $\theta_c$ to zero and therefore connects the finite-temperature non-interacting reference directly to the zero-temperature interacting energy, where $\theta_c\equiv\theta(\lambda=0)$ is the characteristic temperature of the auxiliary non-interacting system. Along this path, the Hamiltonian is
\begin{equation}
\hat{H}^\lambda
=
\hat{T}
+
\hat{v}_{\mathrm{aux}}^\lambda
+
\lambda \hat{V}_{ee},
\label{eq:H_lambda}
\end{equation}
where the auxiliary external potential $\hat v_{\mathrm{aux}}^\lambda$ is adjusted so that the prescribed 1-RDM $\gamma$ is maintained. For each $\lambda\in[0,1]$, we define
\begin{equation}
F^{\lambda}[\gamma]
=
\min_{\hat\Gamma\to\gamma}
\left\{
\mathrm{Tr}\left[
\hat\Gamma
\left(
\hat T+\lambda\hat V_{ee}
\right)
\right]
-
\theta(\lambda) S[\hat\Gamma]
\right\},
\end{equation}
where $\hat\Gamma$ is a statistical density operator constrained to yield the prescribed 1-RDM $\gamma$, $S[\hat\Gamma]=-\mathrm{Tr}(\hat\Gamma\ln\hat\Gamma)$, and the temperature path satisfies $\theta(0)=\theta_c$ and $\theta(1)=0$. The constrained search is performed at fixed $\gamma$ for every value of $\lambda$. Note that the full 1-RDM $\gamma$ is held fixed along the path, in contrast to the density-based adiabatic connection of Kohn--Sham DFT, in which only the density $\rho$ is kept fixed. The auxiliary potential $\hat v_{\mathrm{aux}}^\lambda$ in Eq.~\eqref{eq:H_lambda} acts only as the constraint that maintains this 1-RDM and is not included in $F^\lambda[\gamma]$. This constrained-search construction is equivalent to integrating along the path while retaining the auxiliary potential, as in the grand-potential functional $\Omega_0[\gamma]$ of Eq.~\eqref{eq:omega0}: since $\hat v_{\mathrm{aux}}^\lambda$ is a one-body operator and $\gamma$ is held fixed along the path, its contribution reduces to the endpoint difference $\mathrm{Tr}[\hat v_{\mathrm{aux}}^{\lambda=1}\gamma]-\mathrm{Tr}[\hat v_{\mathrm{aux}}^{\lambda=0}\gamma]$, which cancels when the interacting energy is expressed relative to the non-interacting reference. Excluding the auxiliary potential from $F^\lambda[\gamma]$ from the outset, as done here, avoids this term and makes the coupling-constant integration involve the interaction alone.
The non-interacting endpoint is
\begin{equation}
F^{\lambda=0}[\gamma]
=
T_0[\gamma]-\theta_c S_0[\gamma],
\end{equation}
corresponding to the auxiliary finite-temperature non-interacting system, with $T_0[\gamma]$ and $S_0[\gamma]$ as defined in Eqs.~\eqref{eq:T0}--\eqref{eq:S0} and $\theta_c$ the endpoint temperature of the path.
The endpoint $F^{\lambda=1}[\gamma]$ corresponds to the zero-temperature fully interacting universal functional. The explicit entropy contribution vanishes at this endpoint because $\theta(1)=0$.

Following the definition above, $F^\lambda[\gamma]$ depends on both the coupling strength $\lambda$ and the temperature $\theta(\lambda)$. Its total derivative along the path is therefore
\begin{equation}
\frac{dF^\lambda[\gamma]}{d\lambda}
=
\left.
\frac{\partial F^\lambda}{\partial\lambda}
\right|_{\theta}
+
\left.
\frac{\partial F^\lambda}{\partial \theta}
\right|_{\lambda}
\frac{d\theta}{d\lambda}.
\end{equation}
The first term follows from the Hellmann--Feynman (envelope) theorem. Because $F^\lambda[\gamma]$ is defined by minimizing over ensembles that yield $\gamma$, the minimizing ensemble is variationally stationary, so its change with $\lambda$ does not contribute to the derivative; only the explicit $\lambda$ dependence of $\hat T+\lambda\hat V_{ee}$ remains, yielding simply $\langle\hat V_{ee}\rangle$. Thus one obtains
\begin{equation}
\left.
\frac{\partial F^\lambda}{\partial\lambda}
\right|_{\theta}
=
\left\langle
\frac{\partial( \hat{T} + \lambda \hat{V}_{ee} )}
{\partial\lambda}
\right\rangle_{\lambda,\theta(\lambda),\gamma}
=
\langle\hat V_{ee}\rangle_{\lambda,\gamma}.
\end{equation}
Here and below, $\langle\hat V_{ee}\rangle_{\lambda,\gamma}$ denotes the expectation value in the constrained minimizing ensemble at coupling strength $\lambda$, temperature $\theta(\lambda)$, and fixed 1-RDM $\gamma$.
The second term follows from the thermodynamic relation
\begin{equation}
\left.
\frac{\partial F^\lambda}{\partial \theta}
\right|_{\lambda}
=
-S^\lambda,
\end{equation}
where $S^\lambda=-\mathrm{Tr}\,(\hat\Gamma^\lambda\ln\hat\Gamma^\lambda)$ is the entropy of the constrained minimizing ensemble $\hat\Gamma^\lambda$ at coupling strength $\lambda$, temperature $\theta(\lambda)$, and fixed 1-RDM $\gamma$, i.e.\ the same ensemble that defines $\langle\hat V_{ee}\rangle_{\lambda,\gamma}$ above. Thus
\begin{equation}
\frac{dF^\lambda[\gamma]}{d\lambda}
=
\langle\hat V_{ee}\rangle_{\lambda,\gamma}
-
S^\lambda
\frac{d\theta}{d\lambda}.
\end{equation}
Integrating from $\lambda=0$ to $\lambda=1$ gives
\begin{equation}
F^{\lambda=1}[\gamma]
=
T_0[\gamma]-\theta_c S_0[\gamma]
+
\int_0^1
\langle\hat V_{ee}\rangle_{\lambda,\gamma}
d\lambda
-
\int_0^1
S^\lambda
\frac{d\theta}{d\lambda}
d\lambda .
\end{equation}
Consequently, the physical zero-temperature energy is
\begin{equation}
\begin{aligned}
E[\gamma]
&=
T_0[\gamma]-\theta_c S_0[\gamma]
+
E_{\mathrm{ext}}[\gamma]
\\
&\quad
+
\int_0^1
\langle\hat V_{ee}\rangle_{\lambda,\gamma}
d\lambda
-
\int_0^1
S^\lambda
\frac{d\theta}{d\lambda}
d\lambda .
\end{aligned}
\label{eq:total_ac_energy}
\end{equation}
This relation is the desired adiabatic-connection construction: it connects the finite-temperature non-interacting reference contribution to the zero-temperature interacting energy through an interaction integral and an entropy contribution from the temperature path.
The physical external-potential contribution $E_{\mathrm{ext}}[\gamma]$ is reinstated only at this stage through $E[\gamma]=F^{\lambda=1}[\gamma]+E_{\mathrm{ext}}[\gamma]$; it is the physical external potential of the original system and should not be confused with the auxiliary potential used to enforce the fixed 1-RDM along the path.

The exchange-correlation contribution generated along the adiabatic path is defined by
\begin{equation}
F_{\mathrm{xc}}^{\mathrm{AC}}[\gamma]
=
F^{\lambda=1}[\gamma]
-
F^{\lambda=0}[\gamma]
-
E_{\mathrm{H}}[\gamma].
\end{equation}
Substituting the integrated adiabatic-connection expression gives
\begin{equation}
F_{\mathrm{xc}}^{\mathrm{AC}}[\gamma]
=
\int_0^1
\left(
\langle\hat V_{ee}\rangle_{\lambda,\gamma}
-
E_{\mathrm{H}}[\gamma]
\right)
d\lambda
-
\int_0^1
S^\lambda
\frac{d\theta}{d\lambda}
d\lambda .
\label{eq:F_xc_AC}
\end{equation}
At the non-interacting end of the path, the expectation value of $\hat V_{ee}$ is not zero; for a fermionic non-interacting reference it is the Hartree plus exact-exchange contribution,
\begin{equation}
\langle\hat V_{ee}\rangle_{0,\gamma}
=
E_{\mathrm{H}}[\gamma]+E_{\mathrm{x}}^{\mathrm{EX}}[\gamma].
\label{eq:vee_endpoint}
\end{equation}

We now evaluate the two integrals in Eq.~\eqref{eq:F_xc_AC} to leading order. We adopt a linear temperature path,
\begin{equation}
\theta(\lambda)=\theta_c(1-\lambda),
\qquad
\frac{d\theta}{d\lambda}=-\theta_c,
\end{equation}
which is a free choice of the connection path, the auxiliary potential $\hat v_{\mathrm{aux}}^\lambda$ being adjusted at each $\lambda$ to keep $\gamma$ fixed. By ``first order'' we mean that each of the two integrands in Eq.~\eqref{eq:F_xc_AC} is replaced by its leading behavior along the path: the interaction integrand $\langle\hat V_{ee}\rangle_{\lambda,\gamma}$ is approximated by its value at the non-interacting endpoint (Eq.~\eqref{eq:vee_endpoint}), thereby retaining Hartree and exact exchange but neglecting the coupling-strength dependence that would generate correlation beyond exchange, while the constrained entropy $S^\lambda$ is interpolated linearly from the finite-temperature reference at $\lambda=0$ to zero at the zero-temperature endpoint $\lambda=1$,
\begin{equation}
\langle\hat V_{ee}\rangle_{\lambda,\gamma}
\approx
E_{\mathrm{H}}[\gamma]+E_{\mathrm{x}}^{\mathrm{EX}}[\gamma],
\qquad
S^\lambda\approx S_0[\gamma](1-\lambda).
\label{eq:1st-order_approx}
\end{equation}
With the linear temperature path, these lead to
\begin{equation}
F_{\mathrm{xc}}^{\mathrm{AC}}[\gamma]
\approx
E_{\mathrm{x}}^{\mathrm{EX}}[\gamma]
+
\frac{\theta_c}{2}S_0[\gamma],
\label{eq:F_xc_AC_appro}
\end{equation}
where the factor $1/2$ arises from $\int_0^1 (1-\lambda)\,d\lambda=1/2$.
Combining Eqs.~\eqref{eq:total_ac_energy}, \eqref{eq:F_xc_AC}, and \eqref{eq:F_xc_AC_appro}, the first-order adiabatic-connection approximation gives the total energy
\begin{equation}
E[\gamma]
\approx
E_{\mathrm{HF}}[\gamma]
-
\frac{\theta_c}{2}S_0[\gamma].
\end{equation}
Comparing this expression with the standard i-DMFT form
$E_{\mathrm{i\mbox{-}DMFT}}
=
E_{\mathrm{HF}}
-
\kappa S
-
b,$
where $b$ is a constant energy shift, we identify
\begin{equation}
\kappa
=
\frac{\theta_c}{2}.
\label{eq:kappa_relation}
\end{equation}
Equation~\eqref{eq:kappa_relation} is a key result. In i-DMFT, $\kappa$ is an empirical parameter without a first-principles derivation: it is determined by fitting, either to the dissociation energy or to the entropy extracted from a wave-function calculation~\cite{wang2022scf,irimia2023self,wang2021information}, and it has been noted only heuristically to play the role of an effective temperature~\cite{irimia2023self}. More generally, the use of an entropy of the occupation numbers as an approximation to the correlation energy has recently been placed on a more formal footing~\cite{martinez2023entropy}; there, however, the correlation energy is assumed to be a functional of the occupation numbers alone, and the associated parameter enters merely as the width of an (invertible) distribution rather than as a physical temperature, so that the resulting entropy coefficient remains semi-empirical, as in i-DMFT. The present adiabatic connection instead relates $\kappa$ directly to a physical quantity, the characteristic temperature $\theta_c$ of the auxiliary non-interacting system, i.e.\ the temperature at which that system reproduces the 1-RDM of the interacting system.

This identification should be understood within the present first-order, linear-temperature-path approximation: different choices of the path $\theta(\lambda)$, or higher-order treatments of the coupling-constant integrand, would in general modify the numerical prefactor. The factor of $1/2$ is thus specific to the linear path adopted here. This path dependence is, in fact, consistent with the empirical behavior of i-DMFT, in which the optimal $\kappa$ is found to be system-dependent, and in some cases even orbital-dependent~\cite{liu2024potential}: a single universal value is not expected once the temperature enters through a variable connection path. The clean identification $\kappa=\theta_c/2$ should therefore be read as a first-order relation that supplies the physical origin and scale of the i-DMFT entropy coefficient, rather than as a strict system-independent identity.

This adiabatic-connection interpretation also clarifies the origin and the approximation level of the entropy term in i-DMFT. The final contribution proportional to $S_0[\gamma]$ results from the combination of the finite-temperature non-interacting reference term and the entropy accumulated along the temperature-dependent adiabatic path. The linear path $\theta(\lambda)=\theta_c(1-\lambda)$ and the first-order treatment of the interaction integrand therefore constitute the simplest realization of this construction. More flexible temperature paths or higher-order approximations to the coupling-constant integrand would provide a systematic route beyond the present first-order connection.

\begingroup

\subsection{Energy decomposition and reduction to TAO-DFT}\label{sec:decomposition}

The adiabatic connection of the previous subsection gives a physical meaning to the i-DMFT entropy coefficient. We now turn to the other reduction, the connection to density-based formulations such as TAO-DFT. To this end we recast the universal functional $F[\gamma]$ by adding and subtracting the non-interacting kinetic free energy,
\begin{equation}
T_0[\gamma]-\theta S_0[\gamma].
\end{equation}
To make contact with density-functional theory, we further adopt the Kohn--Sham decomposition of the kinetic energy. For the $v$-representable case considered here, in which $\gamma$ and $\rho$ derive from the same physical system,
\begin{equation}
T[\gamma] = T[\rho] = T_s[\rho] + T_c[\rho],
\qquad
U_c[\gamma] = U_c[\rho],
\end{equation}
where $\rho(\mathbf{r})=\gamma(\mathbf{r},\mathbf{r})$, $T_s[\rho]$ is the zero-temperature Kohn--Sham kinetic energy, and $T_c[\rho]$ the kinetic correlation energy. This partition is not needed within RDMFT itself, where $T[\gamma]$ is already exact; it is introduced only to bring in the Kohn--Sham reference $T_s[\rho]$ used by TAO-DFT. We stress that the $v$-representability assumption is used only to motivate this decomposition and the definitions of $E_c^{\mathrm{DFT}}$ and $E_{\theta}$ below; in practice these terms are represented by approximate functionals defined for any 1-RDM, and the variational minimization of Sec.~\ref{sec:variational} is carried out over the broader set of ensemble-$N$-representable $\gamma$.

Starting from $F[\gamma]=T[\gamma]+V_{ee}[\gamma]$ and separating the exact-exchange contribution, the reorganized functional reads
\begin{equation}
F[\gamma]
= T_0[\gamma]-\theta S_0[\gamma]
+ E_{\mathrm{H}}[\gamma]
+ E_{\mathrm{x}}^{\mathrm{EX}}[\gamma]
+ E_c^{\mathrm{DFT}}[\rho]
+ E_{\theta}[\gamma],
\label{eq:F_decomp}
\end{equation}
with $E_c^{\mathrm{DFT}}[\rho] \equiv T_c[\rho] + U_c[\rho]$ and $E_{\theta}[\gamma] \equiv T_s[\rho] - \big( T_0[\gamma]-\theta S_0[\gamma] \big)$.

The term $E_{\theta}$ reflects what the non-interacting reference is required to reproduce. When it reproduces the full 1-RDM $\gamma$, its kinetic energy equals the interacting one, $T_0[\gamma]=T[\gamma]$, so that, using the Kohn--Sham partition, $E_{\theta}=\theta S_0[\gamma]-T_c[\rho]$; when it reproduces only the density $\rho$, as in TAO-DFT, the relevant reference is the zero-temperature Kohn--Sham kinetic energy $T_s[\rho]$. The term $E_{\theta}$ is simply the difference between these two kinetic references.

The decomposition~\eqref{eq:F_decomp} contains i-DMFT and TAO-DFT as two reductions. i-DMFT corresponds to the exact-exchange-based reduction along the main FT-AC-RDMFT line: the exchange-correlation contribution is the exact exchange $E_{\mathrm{x}}^{\mathrm{EX}}[\gamma]$ supplemented by the entropic term $-\theta S_0[\gamma]$, whose coefficient is fixed by the adiabatic connection of Sec.~\ref{sec:adiabatic} ($\kappa=\theta_c/2$ at first order). TAO-DFT corresponds to the density-based reduction, in which the density-based exchange-correlation functional $E_{\mathrm{xc}}^{\mathrm{DFT}}[\rho]$ is employed, following the spirit of Kohn--Sham DFT, and $E_{\theta}$ is retained to account for the change of kinetic reference. The choice of exchange is not rigid: hybrid schemes that combine exact and density-based exchange within TAO-DFT have also been developed~\cite{chai2017role}.

Although the two reductions differ in what the non-interacting reference is required to reproduce---the full 1-RDM $\gamma$ or only the density $\rho$---this is a distinction at the level of the formal decomposition. The resulting working equations, derived in Sec.~\ref{sec:variational}, are largely the same self-consistent single-particle problem, and in practice neither the exact $\gamma$ nor the exact $\rho$ is obtained; the actual performance is therefore governed by the approximate functionals employed rather than by this formal distinction. This also means that different combinations of exchange, density-based correlation, and $E_{\theta}$ can be explored within the same set of working equations. Indeed, since the exact $\gamma$ is not strictly reproduced once approximations are made, $T_0[\gamma]$ is itself no longer exact, which leaves freedom in how the kinetic and exchange-correlation contributions are corrected.

These relationships are summarized in Table~\ref{tab:reductions}, which classifies the methods by three ingredients of the decomposition: the exchange treatment, the presence of a density-based correlation functional $E_c^{\mathrm{DFT}}$, and the presence of $E_{\theta}$. Besides i-DMFT and TAO-DFT, the table includes FT-HF (exact exchange with neither $E_c^{\mathrm{DFT}}$ nor $E_{\theta}$), FT-HF-$\theta$ (exact exchange supplemented by a density-based correlation functional and $E_{\theta}$), and hybrid TAO-DFT (hybrid exchange); FT-HF and FT-HF-$\theta$ are examined numerically in Sec.~\ref{sec:results}. Here $E_c^{\mathrm{DFT}}$ refers specifically to a density-based correlation functional, while the entropy term $-\theta S_0$ is listed separately; the entry ``none (via entropy)'' for i-DMFT indicates that correlation is carried by the entropy term rather than by a density-based functional, which is the same physical content as the original i-DMFT description of the entropy as a (cumulant) correlation energy. The temperature $\theta$ plays the role of the empirical coefficient $\kappa$ in i-DMFT (with $\kappa=\theta_c/2$ at first order) and of the fictitious temperature in TAO-DFT, so that a single parameter unifies these otherwise distinct quantities.

\begin{table}[t]
\caption{Methods obtained as reductions of FT-AC-RDMFT for different choices of the three ingredients of the energy decomposition: the exchange treatment, the density-based correlation functional $E_c^{\mathrm{DFT}}$, and the finite-temperature correction $E_{\theta}$. Here $E_{\mathrm{x}}^{\mathrm{EX}}$ denotes exact exchange and $E_{\mathrm{xc}}^{\mathrm{DFT}}$ a density-based exchange-correlation functional (in which exchange and correlation are approximated together). The entry ``none (via entropy)'' indicates that correlation is carried by the entropy term $-\theta S_0$ rather than by a density-based correlation functional (see text).}
\label{tab:reductions}
\begin{ruledtabular}
\begin{tabular}{lccc}
Method & Exchange & $E_c^{\mathrm{DFT}}$ & $E_{\theta}$ \\
\colrule
i-DMFT & exact ($E_{\mathrm{x}}^{\mathrm{EX}}$) & none (via entropy) & no \\
TAO-DFT & density-based & yes ($E_{\mathrm{xc}}^{\mathrm{DFT}}$) & yes \\
TAO-DFT-hybrid & hybrid & yes ($E_{\mathrm{xc}}^{\mathrm{DFT}}$) & yes \\
FT-HF & exact ($E_{\mathrm{x}}^{\mathrm{EX}}$) & no & no \\
FT-HF-$\theta$ & exact ($E_{\mathrm{x}}^{\mathrm{EX}}$) & yes & yes \\
\end{tabular}
\end{ruledtabular}
\end{table}

\subsection{Variational equations and Fermi--Dirac occupations}\label{sec:variational}

To determine the ground state, we minimize the total energy of Eq.~\eqref{eq:F_decomp} together with the external-potential term,
\begin{equation}
\begin{aligned}
E[\gamma]
&=
T_0[\gamma] - \theta S_0[\gamma]
+ E_{\mathrm{ext}}[\gamma]
+ E_{\mathrm{H}}[\gamma]
\\
&\quad
+ E_{\mathrm{x}}^{\mathrm{EX}}[\gamma]
+ E_c^{\mathrm{DFT}}[\rho]
+ E_{\theta}[\gamma],
\end{aligned}
\label{eq:E_total}
\end{equation}
with $\rho(\mathbf r)=\gamma(\mathbf r,\mathbf r)$, under the ensemble-$N$-representability constraints of the 1-RDM. For the local external potential considered here, $E_{\mathrm{ext}}[\gamma]=\int \rho(\mathbf r)\,v_{\mathrm{ext}}(\mathbf r)\,d\mathbf r$ depends on $\gamma$ only through $\rho$. The corresponding Lagrangian is
\begin{equation}
\Omega[\gamma]
=
E[\gamma]
-
\sum_{ij}
\varepsilon_{ij}
\big(
\langle \varphi_i | \varphi_j \rangle
-
\delta_{ij}
\big)
-
\mu
\left(
\sum_i n_i - N
\right),
\end{equation}
where $\varepsilon_{ij}$ and $\mu$ are Lagrange multipliers associated with the orthonormality of the natural orbitals and the particle-number constraint, respectively. We first consider the variation with respect to the natural orbitals,
\begin{equation}
\frac{\delta \Omega}{\delta \varphi_j^*(\mathbf r)} = 0.
\end{equation}

Since $E_{\mathrm{ext}}$, $E_{\mathrm{H}}$, and $E_c^{\mathrm{DFT}}$ depend on $\gamma$ only through the density $\rho(\mathbf r)$, their orbital derivatives follow from the chain rule with $\delta\rho(\mathbf r_1)/\delta\varphi_j^*(\mathbf r)=\delta(\mathbf r-\mathbf r_1)\,n_j\varphi_j(\mathbf r_1)$, giving local potentials $v_{\mathrm{ext}}(\mathbf r)=\delta E_{\mathrm{ext}}/\delta\rho$, $v_{\mathrm{H}}(\mathbf r)=\delta E_{\mathrm{H}}/\delta\rho$, and $v_c^{\mathrm{DFT}}(\mathbf r)=\delta E_c^{\mathrm{DFT}}/\delta\rho$. The remaining two terms, the exact exchange $E_{\mathrm{x}}^{\mathrm{EX}}[\gamma]$ and $E_{\theta}[\gamma]$, define the operators $\hat v_{\mathrm{x}}^{\mathrm{EX}}$ and $\hat v_{\theta}$ through
\begin{equation}
\frac{\delta E_{\mathrm{x}}^{\mathrm{EX}}[\gamma]}{\delta \varphi_j^*(\mathbf r)}
=
n_j \hat v_{\mathrm{x}}^{\mathrm{EX}} \varphi_j(\mathbf r),
\qquad
\frac{\delta E_{\theta}[\gamma]}{\delta \varphi_j^*(\mathbf r)}
=
n_j \hat v_{\theta} \varphi_j(\mathbf r).
\end{equation}
For the exact-exchange term,
\begin{equation}
\hat v_{\mathrm{x}}^{\mathrm{EX}} \varphi_j(\mathbf r)
=
-\sum_i n_i \hat K_i \varphi_j(\mathbf r),
\end{equation}
where $\hat K_i$ is the usual Hartree--Fock exchange operator. The stationary condition for the natural orbitals then becomes
\begin{equation}
\begin{aligned}
n_j
\Big[
-\frac{1}{2}\nabla^2
+
v_{\mathrm{ext}}(\mathbf r)
+
v_{\mathrm{H}}(\mathbf r)
+
v_c^{\mathrm{DFT}}(\mathbf r)
\Big]
\varphi_j(\mathbf r)
\\
\qquad
+
n_j
\big(
\hat v_{\mathrm{x}}^{\mathrm{EX}}
+
\hat v_{\theta}
\big)
\varphi_j(\mathbf r)
-
\sum_k \varepsilon_{jk}\varphi_k(\mathbf r)
=
0.
\end{aligned}
\end{equation}

It is convenient to define the effective one-particle operator
\begin{equation}
\hat h
=
-\frac{1}{2}\nabla^2
+
v_{\mathrm{ext}}(\mathbf r)
+
v_{\mathrm{H}}(\mathbf r)
+
v_c^{\mathrm{DFT}}(\mathbf r)
+
\hat v_{\mathrm{x}}^{\mathrm{EX}}
+
\hat v_{\theta}.
\end{equation}
The Euler equation for the natural orbitals can then be written as
\begin{equation}
n_j \hat h |\varphi_j\rangle
=
\sum_k \varepsilon_{jk} |\varphi_k\rangle.
\end{equation}
Projecting onto $\langle \varphi_i|$ gives
$n_j \langle \varphi_i|\hat h|\varphi_j\rangle = \varepsilon_{ji}.$
Since $\varepsilon_{ij}=\varepsilon_{ji}^*$ and $\hat h$ is Hermitian, one finds
\begin{equation}
(n_j-n_i)\langle \varphi_i|\hat h|\varphi_j\rangle = 0.
\end{equation}
Therefore, the off-diagonal matrix elements of $\hat h$ vanish for orbitals with different occupation numbers. Within each subspace of orbitals sharing the same occupation number, the remaining unitary freedom can be used to diagonalize $\hat h$. We note that this step corresponds to the degeneracy difficulty encountered in zero-temperature RDMFT, where fractionally occupied orbitals become effectively degenerate and an eigenvalue equation of Kohn--Sham type cannot in general be defined. Here the finite-temperature entropy term lifts this degeneracy through the Fermi--Dirac relation between occupation numbers and orbital energies, so that a well-defined single-particle eigenvalue equation can always be imposed. It follows that the natural orbitals can be chosen to satisfy an eigenvalue equation of the form
\begin{equation}
\hat h |\varphi_i\rangle
=
\varepsilon_i |\varphi_i\rangle.
\label{eq:effective_eigen}
\end{equation}

We next consider the variation with respect to the occupation numbers,
\begin{equation}
\frac{\partial \Omega}{\partial n_j}=0.
\end{equation}
Since
\begin{equation}
\frac{\partial \rho(\mathbf r)}{\partial n_j}
=
|\varphi_j(\mathbf r)|^2,
\end{equation}
the stationarity condition can be written as
\begin{equation}
\begin{aligned}
\frac{\partial \Omega}{\partial n_j}
&=
-\frac{1}{2}
\int
\varphi_j^*(\mathbf r)\nabla^2\varphi_j(\mathbf r)\,d\mathbf r
\\
&\quad
+
\int
\frac{\delta \big(E_{\mathrm{ext}}+E_{\mathrm{H}}+E_c^{\mathrm{DFT}}\big)}
{\delta \rho(\mathbf r)}
|\varphi_j(\mathbf r)|^2\,d\mathbf r
\\
&\quad
+
\frac{\partial E_{\mathrm{x}}^{\mathrm{EX}}[\gamma]}{\partial n_j}
+
\frac{\partial E_{\theta}[\gamma]}{\partial n_j}
+
\theta \ln\frac{n_j}{1-n_j}
-
\mu
=
0.
\end{aligned}
\end{equation}
Using
\begin{equation}
\begin{gathered}
\begin{aligned}
\frac{\partial E_{\mathrm{x}}^{\mathrm{EX}}[\gamma]}{\partial n_j}
&=
\langle \varphi_j | \hat v_{\mathrm{x}}^{\mathrm{EX}} | \varphi_j \rangle,
&
\frac{\partial E_{\theta}[\gamma]}{\partial n_j}
&=
\langle \varphi_j | \hat v_{\theta} | \varphi_j \rangle
\end{aligned}
\end{gathered}
\end{equation}
As in Kohn--Sham DFT, where the exchange-correlation energy $E_{\mathrm{xc}}[\rho]$ has no explicit density dependence but its functional derivative $v_{\mathrm{xc}}=\delta E_{\mathrm{xc}}/\delta\rho$ is well defined and enters the single-particle equations, $E_{\theta}$ need not be given an explicit closed form. The zero-temperature Kohn--Sham kinetic energy $T_s[\rho]$ is not an explicit functional of $\rho$, yet $\delta T_s/\delta\rho$ is well defined; the derivatives $\delta E_{\theta}/\delta\varphi_j^*$ and $\partial E_{\theta}/\partial n_j$ are therefore obtained by the standard chain rule, with $\partial\rho(\mathbf r)/\partial n_j=|\varphi_j(\mathbf r)|^2$, and are packaged into the effective potential $\hat v_{\theta}$. In practice only the difference $E_{\theta}=T_s[\rho]-\big(T_0[\gamma]-\theta S_0[\gamma]\big)$ enters, which is presumably small and is approximated at the local-density level (Sec.~\ref{sec:lda_etheta}).

The variation with respect to $n_j$ reduces to
\begin{equation}
\frac{\partial \Omega}{\partial n_j}
=
\langle \varphi_j | \hat h | \varphi_j \rangle
+
\theta \ln\frac{n_j}{1-n_j}
-
\mu
=
0.
\end{equation}
From the eigenvalue equation for $\hat h$, we have
$\langle \varphi_j | \hat h | \varphi_j \rangle = \varepsilon_j.$
The occupation numbers therefore take the Fermi--Dirac form,
\begin{equation}
n_j
=
\frac{1}{1+\exp\big[(\varepsilon_j-\mu)/\theta\big]}.
\label{eq:fd_occupation}
\end{equation}
In contrast to TAO-DFT, where the Fermi--Dirac distribution is imposed as an ansatz, here it emerges directly from the variational minimization with respect to the occupation numbers.

In summary, the FT-AC-RDMFT formulation leads to a diagonalization-based self-consistent scheme, in which the natural orbitals satisfy the eigenvalue equation~\eqref{eq:effective_eigen} and the occupation numbers follow the Fermi--Dirac form~\eqref{eq:fd_occupation}, with the total energy given by Eq.~\eqref{eq:E_total}. The finite-temperature occupation structure is not imposed externally but arises from the variational formulation in terms of the 1-RDM.
Taken together, these results show how the two reductions identified in Sec.~\ref{sec:decomposition} at the level of the energy functional are realized at the level of the working equations. Retaining the exact-exchange term yields a Fock-type operator with the nonlocal exchange potential $\hat v_{\mathrm{x}}^{\mathrm{EX}}$, as in i-DMFT, whereas replacing it by a density-based exchange-correlation functional yields a purely local Kohn--Sham-type potential, as in TAO-DFT; hybrid treatments, such as hybrid TAO-DFT, interpolate between these limits, giving an effective potential that is partly nonlocal and partly local. The choice of exchange, made at the level of the energy functional, thus determines the degree of nonlocality of the effective single-particle potential, which ranges continuously from fully nonlocal to fully local.

\subsection{Local-density approximation for $E_{\theta}$}\label{sec:lda_etheta}

Since the exact form of $E_{\theta}[\gamma]$ is not known, an approximation is required in practical calculations. In the present work, we adopt a local-density approximation (LDA) for $E_{\theta}[\gamma]$, defined as
\begin{equation}
E_{\theta}[\gamma]
\equiv
T_s[\rho]
-
\big(
T_0[\gamma]-\theta S_0[\gamma]
\big).
\end{equation}
Following the TAO-DFT treatment~\cite{chai2012density}, we approximate the finite-temperature non-interacting term by its LDA form and the zero-temperature kinetic energy by its local-density (Thomas--Fermi) form, namely,
\begin{equation}
T_0[\gamma]-\theta S_0[\gamma]
\approx
A_s^{\mathrm{LDA},\theta}[\rho]
=
\int a_s^{\mathrm{LDA},\theta}(\mathbf r)\, d\mathbf r,
\end{equation}
\begin{equation}
T_s[\rho]
\approx
T_s^{\mathrm{TF}}[\rho]
=
C_F \int \rho^{5/3}(\mathbf r)\, d\mathbf r,
\end{equation}
with $C_F=\frac{3}{10}(3\pi^2)^{2/3}$. Here, $a_s^{\mathrm{LDA},\theta}(\mathbf r)$ is taken from the standard parametrization of the finite-temperature uniform electron gas given in Ref.~\cite{perrot1979gradient}. The resulting LDA expression for $E_{\theta}$ is therefore written as
\begin{equation}
E_{\theta}^{\mathrm{LDA}}[\gamma]
=
T_s^{\mathrm{TF}}[\rho]
-
A_s^{\mathrm{LDA},\theta}[\rho].
\end{equation}
This approximation provides a practical route for incorporating $E_{\theta}$ into the FT-AC-RDMFT framework, while making explicit that the LDA treatment of $E_{\theta}$ used in TAO-DFT can be understood within the present formulation.

\section{Computational Results}
\label{sec:results}

All i-DMFT and finite-temperature calculations reported in this work were implemented and carried out within the ABACUS package~\cite{li2016large,lin2024ab,abacus2025aiera}. The calculations employed numerical atomic orbitals within the LCAO representation and norm-conserving pseudopotentials. Triple-zeta double-polarization (TZDP) basis sets were used for H$_2$, HF, and solid hydrogen, whereas double-zeta polarization (DZP) basis sets were used for N$_2$ and corundum-type V$_2$O$_3$. For the periodic systems, uniform $10\times10\times10$ Monkhorst--Pack meshes were used for solid hydrogen and corundum-type V$_2$O$_3$. The exact-exchange contributions were evaluated within the localized resolution-of-identity (LRI) approximation as implemented in ABACUS through LibRI~\cite{lin2020accuracy,lin2021efficient}. We first verify the formal and numerical equivalence between i-DMFT and finite-temperature Hartree--Fock (FT-HF), then compare FT-HF and FT-PBE0 for molecular dissociation, assess the performance of FT-HF-$\theta$ against higher-level references, and finally examine the finite-temperature framework for solid hydrogen and corundum-type V$_2$O$_3$.

\subsection{Equivalence between i-DMFT and FT-HF}\label{sec:idmft_fthf}

Formally, i-DMFT and FT-HF have essentially the same energy expression and the same self-consistent structure. Both contain the same finite-temperature non-interacting term together with the external, Hartree, and exact-exchange contributions. The one distinction is that the i-DMFT energy expression, $E_{\mathrm{HF}}-\kappa S-b$, contains an additional constant $b$ that is absent in FT-HF. Since $b$ is a constant, it shifts the total energy uniformly and does not affect the shape of the binding-energy curve; once the curves are aligned at their minima, i-DMFT and FT-HF become equivalent. They should therefore yield identical total energies, up to this constant, when the same temperature parameter is used. Here ``the same temperature parameter'' means that the coefficient multiplying the entropy term in i-DMFT is chosen consistently with the temperature entering the FT-HF free-energy functional; the constant shift in the i-DMFT expression changes only the energy zero.

To verify this point numerically, we implemented i-DMFT in ABACUS following the self-consistent i-DMFT framework and benchmarked it against the built-in FT-HF implementation in ABACUS. Figure~\ref{fig:idmft_fthf_compare} compares the dissociation curves of H$_2$, N$_2$, and HF obtained from the two methods. For each system, the temperature parameter $\theta$ was determined by adjusting it until the dissociation energy reproduced the corresponding higher-level quantum-chemical benchmark reported in Refs.~\cite{wang2022scf,liu2025potential}. For all three molecules, the i-DMFT and FT-HF curves are indistinguishable within plotting resolution when the same temperature parameter is used. In particular, for H$_2$, the two curves overlap over the entire dissociation coordinate, from the equilibrium region to the large-distance plateau. The same behavior is observed for N$_2$ and HF. These results provide a direct numerical confirmation of the formal identity between i-DMFT and FT-HF in practical self-consistent calculations. This identity is also expected analytically: the first-order (exchange-only) functional of FT-RDMFT coincides with the FT-HF functional~\cite{baldsiefen2015reduced,baldsiefen2017exchange}.

\begin{figure*}[tbp]
\centering
\subfloat[H$_2$]{%
    \includegraphics[width=0.32\textwidth]{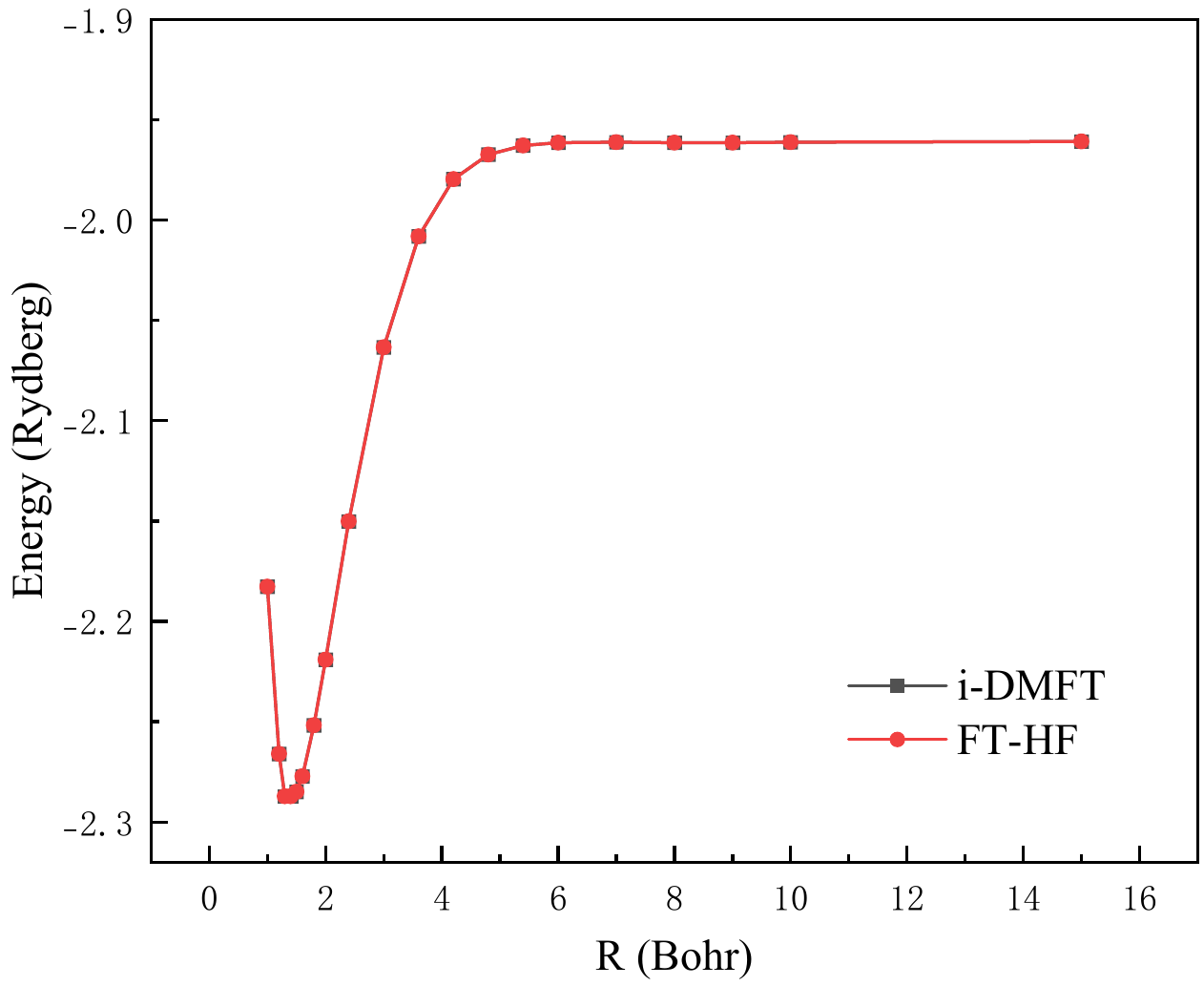}
    \label{fig:idmft_fthf_h2}
}
\hfill
\subfloat[N$_2$]{%
    \includegraphics[width=0.32\textwidth]{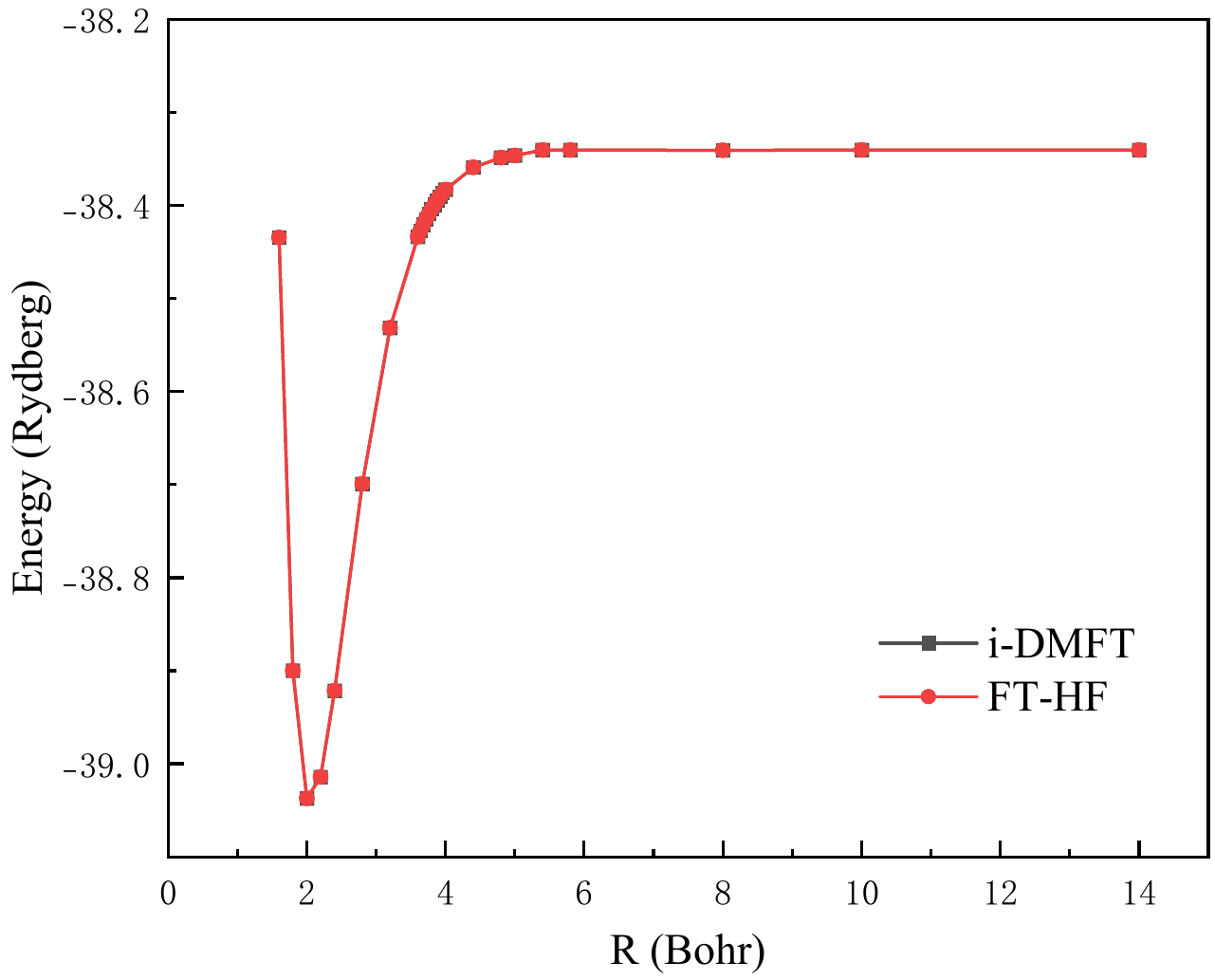}
    \label{fig:idmft_fthf_n2}
}
\hfill
\subfloat[HF]{%
    \includegraphics[width=0.32\textwidth]{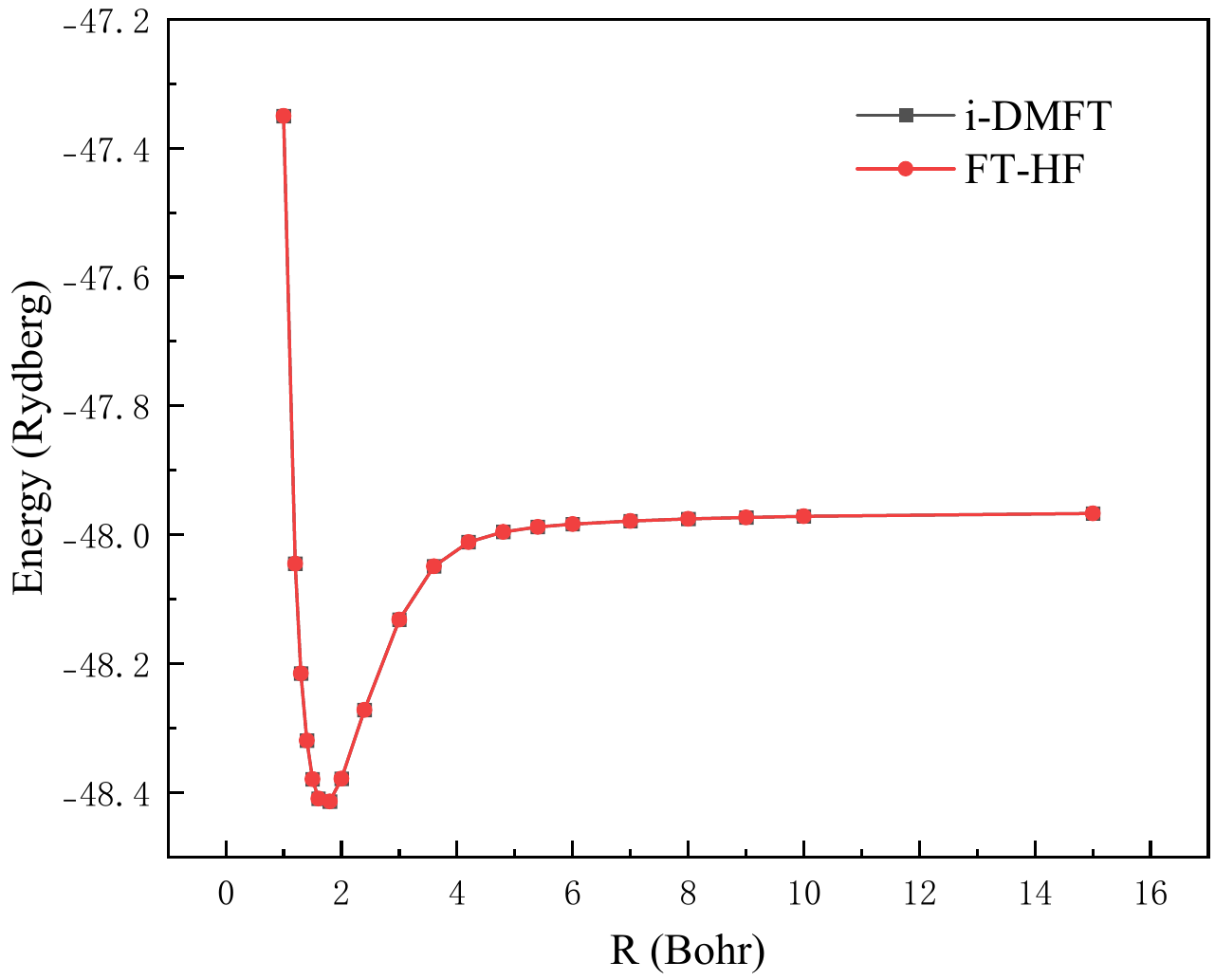}
    \label{fig:idmft_fthf_hf}
}
\caption{
Comparison of the dissociation curves obtained from i-DMFT (with $b=0$) and FT-HF for H$_2$ ($\theta=0.189362$~Ry), N$_2$ ($\theta=0.234$~Ry), and HF ($\theta=0.193$~Ry). For each molecule, the same temperature parameter is used in the two methods. The two curves are indistinguishable within plotting resolution, confirming the numerical equivalence between i-DMFT and FT-HF.
}
\label{fig:idmft_fthf_compare}
\end{figure*}

\subsection{Comparison between FT-HF and FT-PBE0}\label{sec:fthf_ftpbe0}

The favorable dissociation behavior established above raises the question of whether it is tied specifically to exact exchange or reflects a more general effect of the finite-temperature occupations. In this context, we introduce the FT-PBE0 scheme as a means to probe this question. FT-PBE0 denotes the finite-temperature occupation formulation combined with the PBE0 hybrid exchange-correlation functional, without the $E_{\theta}$ correction. Within the classification of Sec.~\ref{sec:decomposition} (Table~\ref{tab:reductions}), it corresponds to hybrid exchange together with a density-based correlation functional but no $E_{\theta}$ term; equivalently, it is obtained from FT-HF by replacing exact exchange with the PBE0 hybrid, or from TAO-DFT-hybrid by omitting $E_{\theta}$. Because FT-HF and FT-PBE0 share the same finite-temperature occupation structure and neither includes $E_{\theta}$, a comparison between them isolates the effect of the exchange treatment alone. We therefore compare the two for the dissociation of N$_2$ and HF, choosing the temperature parameter separately for each functional, since its optimal value is functional dependent.

As shown in Fig.~\ref{fig:fthf_ftpbe0_compare}, FT-HF and FT-PBE0 yield qualitatively the same dissociation behavior for both the N$_2$ and HF molecules, approaching a nearly flat limit after bond breaking. The finite-temperature occupation mechanism is therefore effective not only for exact exchange but also for a hybrid functional such as PBE0, indicating that the improvement in molecular dissociation is not specific to exact exchange and supporting the broader applicability of finite-temperature schemes.

The dissociation profiles are very similar, yet the absolute total energies of FT-HF and FT-PBE0 still differ for a given molecule. Reproducing the correct dissociation behavior by tuning the temperature parameter thus does not by itself guarantee an accurate total energy; the remaining difference reflects the different underlying exchange-correlation approximations.
    
\begin{figure*}[tbp]
\centering
\subfloat[N$_2$]{%
    \includegraphics[width=0.42\textwidth]{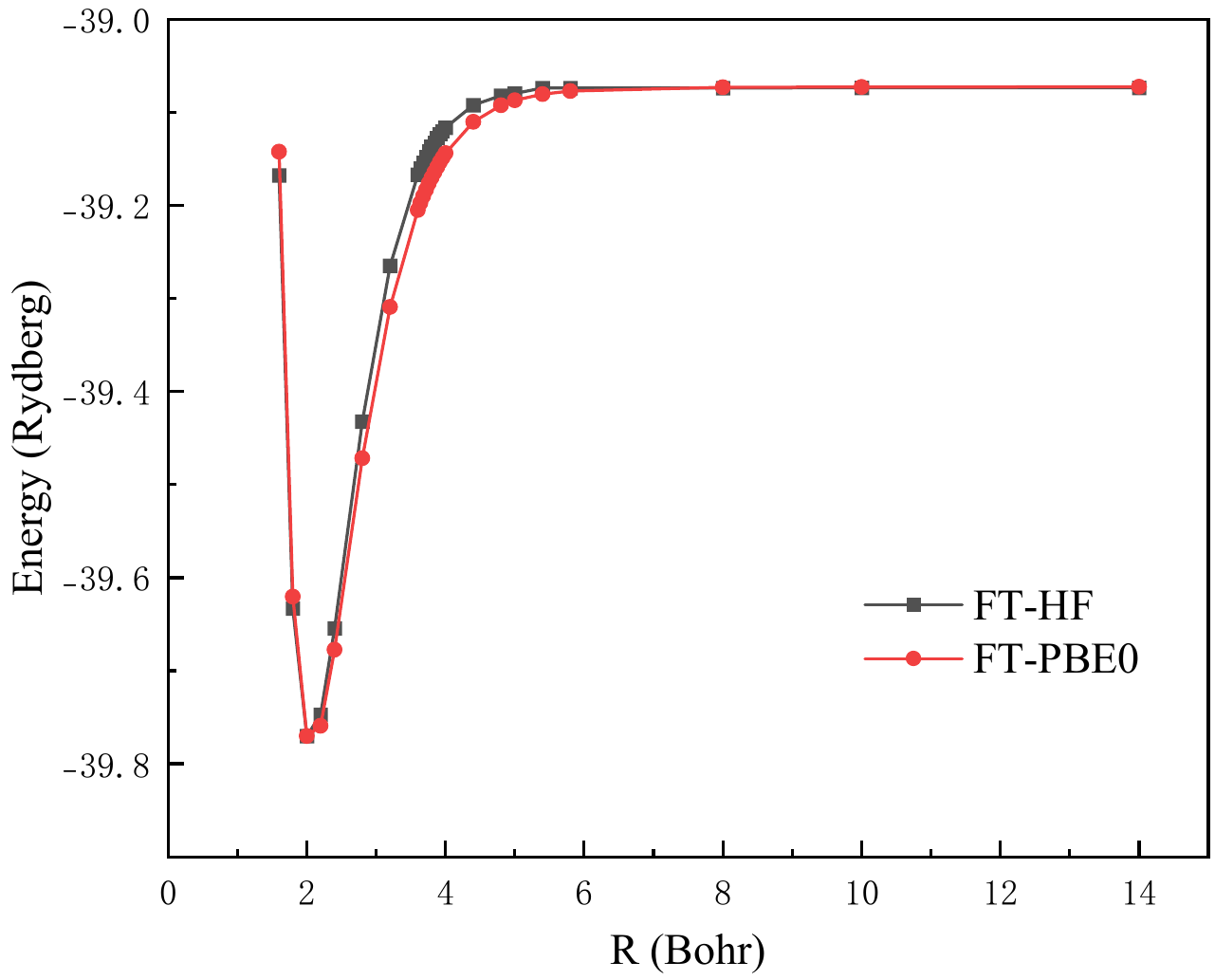}
    \label{fig:fthf_ftpbe0_n2}
}
\hspace{0.03\textwidth}
\subfloat[HF]{%
    \includegraphics[width=0.42\textwidth]{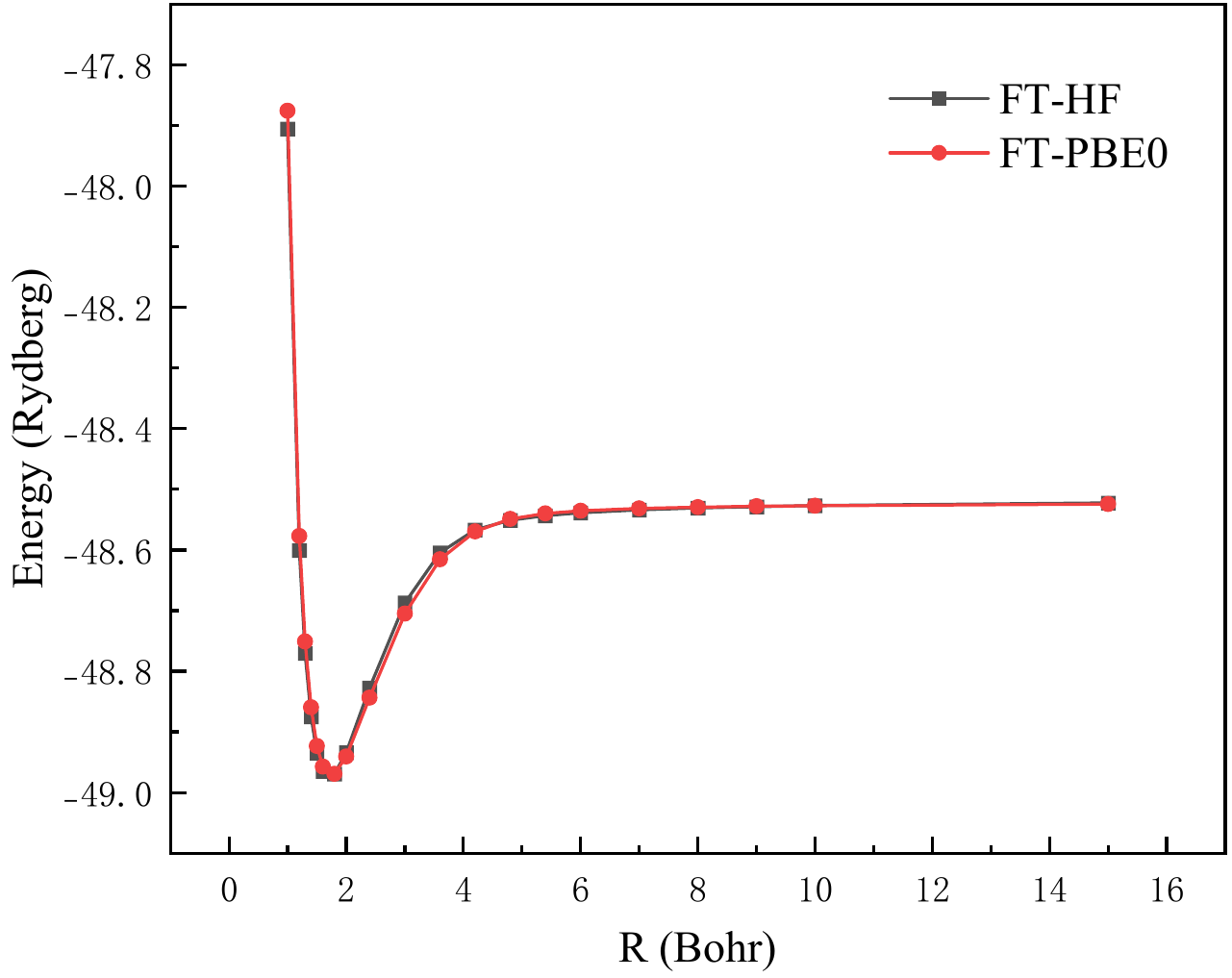}
    \label{fig:fthf_ftpbe0_hf}
}
\caption{
Comparison of the dissociation curves obtained from FT-HF and FT-PBE0 for (a) N$_2$ and (b) HF. The temperature parameters used for FT-HF and FT-PBE0, respectively, are $\theta=0.234$ and $0.1085$~Ry for N$_2$, and $\theta=0.193$ and $0.071$~Ry for HF. After a constant alignment of the absolute energies, the two methods exhibit very similar dissociation profiles, in particular the nearly flat large-distance limit.
}
\label{fig:fthf_ftpbe0_compare}
\end{figure*}

\subsection{Performance of FT-HF-$\theta$ for molecular dissociation}\label{sec:fthf_theta_mol}

The schemes examined so far, FT-HF and FT-PBE0, do not include the $E_{\theta}$ term, and both already capture the qualitative dissociation behavior. We now turn to the full FT-AC-RDMFT scheme that includes $E_{\theta}$, in order to test how well its present local form performs within an exact-exchange-based framework. In the present work, FT-HF-$\theta$ denotes the FT-AC-RDMFT scheme in which the exchange-correlation part is approximated by exact exchange together with LDA correlation; relative to FT-HF (equivalently, i-DMFT), it therefore contains two additional contributions, $E_c^{\mathrm{LDA}}[\rho]$ and $E_{\theta}[\gamma]$ (see Table~\ref{tab:reductions}). Figure~\ref{fig:fthf_theta_mol} compares the dissociation curves of H$_2$, HF, and N$_2$ obtained from FT-HF-$\theta$, FT-HF, and higher-level reference calculations~\cite{wang2022scf,liu2025potential}. For all three molecules, FT-HF-$\theta$ deviates from the reference curves more visibly than FT-HF, particularly in the intermediate bond-stretching region. We have also checked that this trend is unlikely to be dominated by basis-set effects by comparing the present FT-HF results, which are equivalent to i-DMFT at the same temperature, with the i-DMFT reference curves reported in Refs.~\cite{wang2022scf,liu2025potential}. The close agreement between these two i-DMFT/FT-HF descriptions suggests that the larger deviation of FT-HF-$\theta$ from the higher-level references mainly reflects the additional approximate $E_c^{\mathrm{LDA}}+E_{\theta}$ contributions rather than the different basis representations. With their present LDA forms, therefore, the additional $E_c^{\mathrm{LDA}}+E_{\theta}$ contributions do not bring the exact-exchange-based description closer to the reference.

This outcome does not imply that $E_{\theta}$ is unimportant, nor that the framework is at fault: the equivalence in Sec.~\ref{sec:idmft_fthf} and the comparison in Sec.~\ref{sec:fthf_ftpbe0} already show that the finite-temperature occupation structure is effective. Rather, adding the present LDA-based $E_c^{\mathrm{LDA}}+E_{\theta}$ correction on top of exact exchange is not successful in improving the molecular dissociation curves. We note that the same LDA form of $E_{\theta}$ performs reasonably for molecular dissociation within TAO-DFT, where it is combined with a semilocal exchange-correlation functional; its less favorable behavior here may therefore be related to combining a correction designed for a semilocal framework with exact exchange, rather than to a deficiency of $E_{\theta}$ itself.

The role of $E_{\theta}$ in such frameworks is in fact not yet settled. On the one hand, our results show that its present LDA form does not improve the exact-exchange-based description of molecular dissociation; on the other hand, reformulations of TAO-DFT in the Kohn--Sham single-determinant framework reproduce much of the TAO-DFT results without employing the $E_{\theta}$ functional explicitly~\cite{yeh2022reformulation}. Taken together, these observations indicate that both the practical importance of $E_{\theta}$ and the way it is best approximated or incorporated remain open questions, and clarifying them---whether by constructing improved, possibly beyond-LDA, forms of $E_{\theta}$ or by absorbing its effect through a reformulation---is an important direction for future work.

\begin{figure*}[tbp]
\centering
\subfloat[H$_2$]{%
    \includegraphics[width=0.32\textwidth]{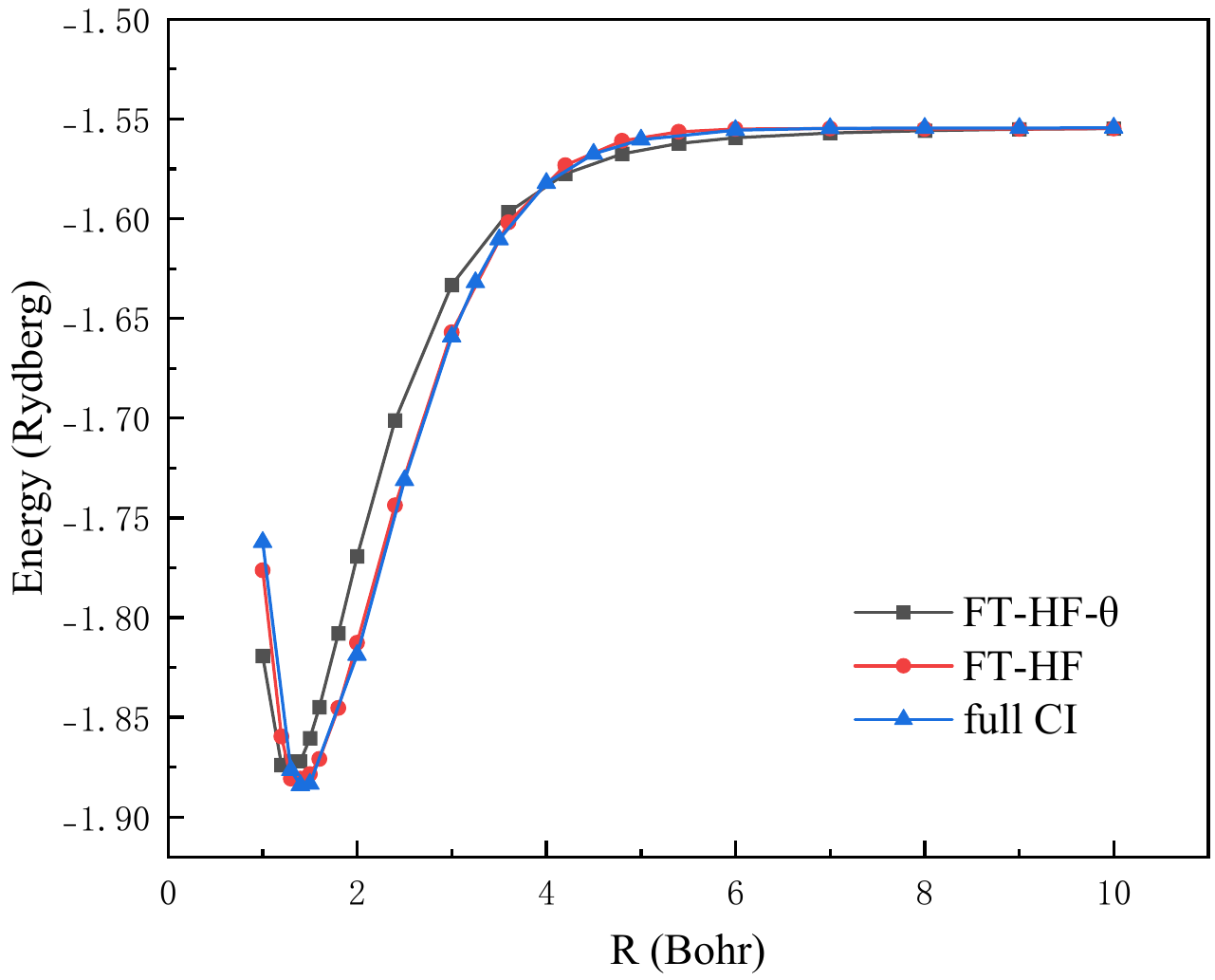}
    \label{fig:fthf_theta_h2}
}
\hfill
\subfloat[HF]{%
    \includegraphics[width=0.32\textwidth]{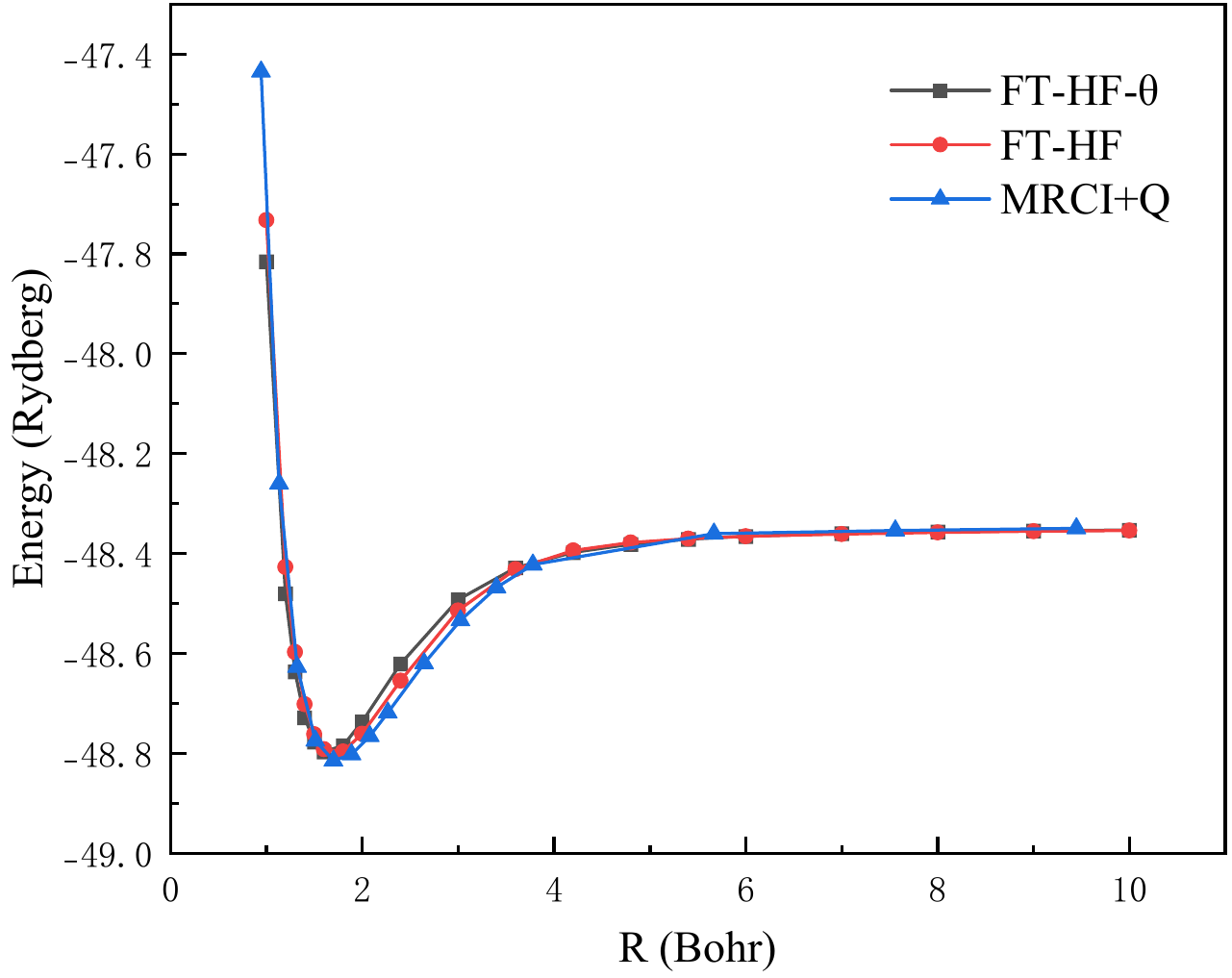}
    \label{fig:fthf_theta_hf}
}
\hfill
\subfloat[N$_2$]{%
    \includegraphics[width=0.32\textwidth]{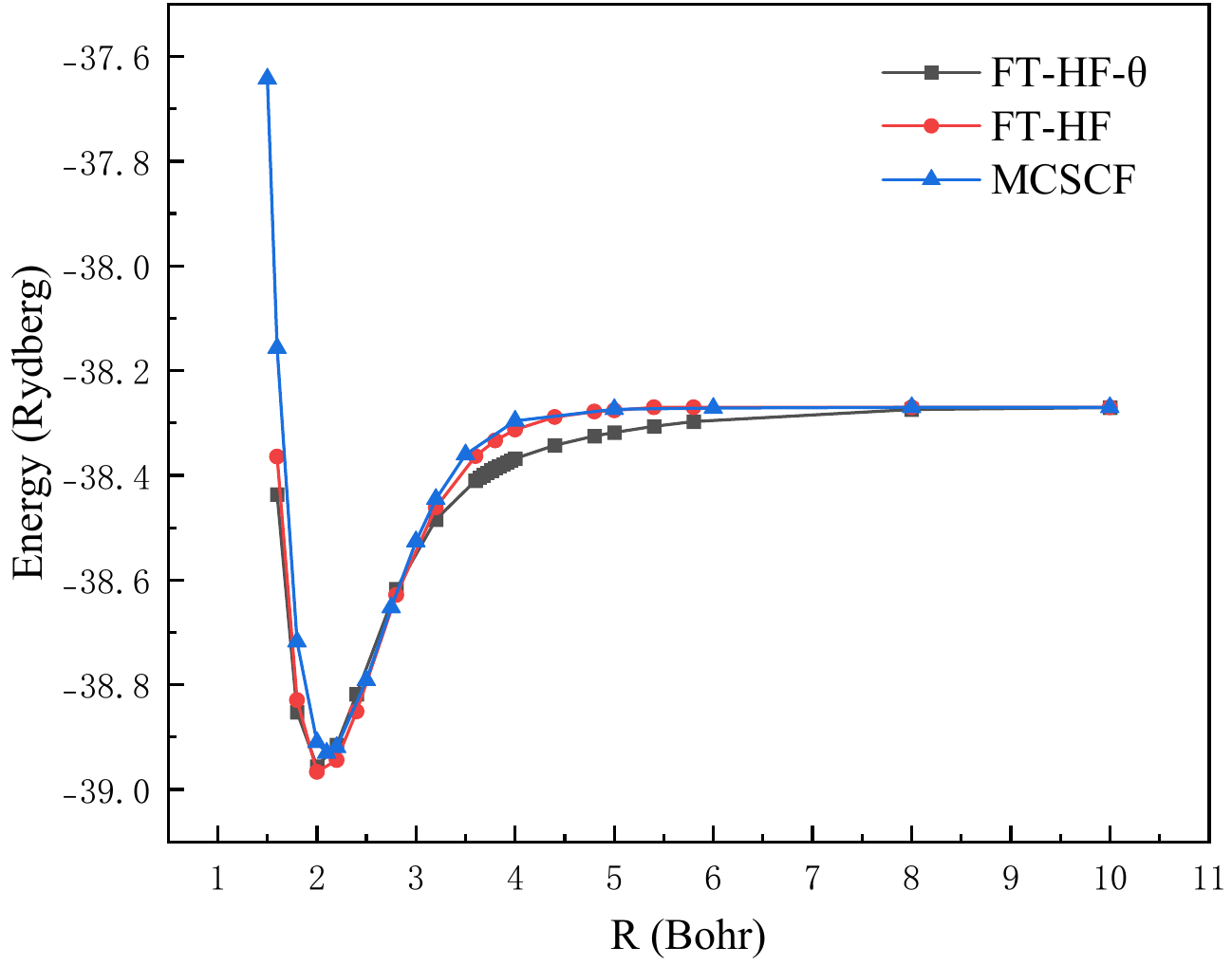}
    \label{fig:fthf_theta_n2}
}
\caption{
Comparison of the dissociation curves obtained from FT-HF-$\theta$, FT-HF (i-DMFT), and higher-level reference calculations for (a) H$_2$ (full CI), (b) HF (MRCI+Q), and (c) N$_2$ (MCSCF). The temperature parameters used in FT-HF-$\theta$ and FT-HF are, respectively, $\theta=0.40$ and $0.18$~Ry for H$_2$, $\theta=0.275$ and $0.193$~Ry for HF, and $\theta=0.32$ and $0.234$~Ry for N$_2$. The three curves are aligned at their dissociation limits for visual comparison. The reference curves for H$_2$ and N$_2$ are taken from Ref.~\cite{wang2022scf}, and the reference curve for HF is taken from Ref.~\cite{liu2025potential}.
}
\label{fig:fthf_theta_mol}
\end{figure*}

\subsection{Dissociation of solid hydrogen}\label{sec:hsolid}

To examine whether the finite-temperature framework remains effective beyond molecular systems, we next consider the dissociation of three-dimensional solid hydrogen as a function of the lattice constant. For comparison, we use the DoNOF-PNOF7 results reported in Ref.~\cite{lew2025advancing}, obtained for the symmetric dissociation of a $10\times10\times10$ cluster of $1000$ hydrogen atoms, which serves as an independent RDMFT description of static correlation. Since the two calculations differ in their boundary conditions---a periodic solid here versus a large finite cluster in the reference---the comparison below is made in terms of the per-atom energy along the dissociation path.

As shown in Fig.~\ref{fig:hsolid_ftpbe_rdmft}, FT-PBE with $\theta=0.122$~Ry agrees well with the RDMFT-PNOF7 per-atom energy curve over the full range of lattice constants considered, including both the minimum region and the large-distance limit. The favorable effect of finite-temperature occupations is therefore not restricted to exact-exchange-based schemes or to isolated molecules: for the present system, a semilocal density functional approximation (DFA) combined with an appropriate temperature parameter recovers the dissociation energetics of the RDMFT (DoNOF-PNOF7) results.

By contrast, the exact-exchange-based schemes deviate more from the DoNOF-PNOF7 curve quantitatively. For FT-HF, the dissociation energy decreases as the temperature is raised, and once $\theta$ exceeds about $0.16$~Ry the large-distance branch bends downward in a physically incorrect manner; for FT-HF-$\theta$, the curves vary only weakly with temperature and lie substantially below the RDMFT benchmark curve. For solid hydrogen, the present exact-exchange-based finite-temperature schemes are thus still insufficient to describe the dissociation energetics quantitatively. This parallels the molecular case of Sec.~\ref{sec:fthf_theta_mol}: the benefit of finite-temperature occupations is robust within semilocal schemes, whereas its combination with exact exchange remains more delicate. The corresponding temperature-dependent curves, including additional FT-HSE results, are provided in the Supplemental Material.

\begin{figure}[t]
\centering
\includegraphics[width=0.95\columnwidth]{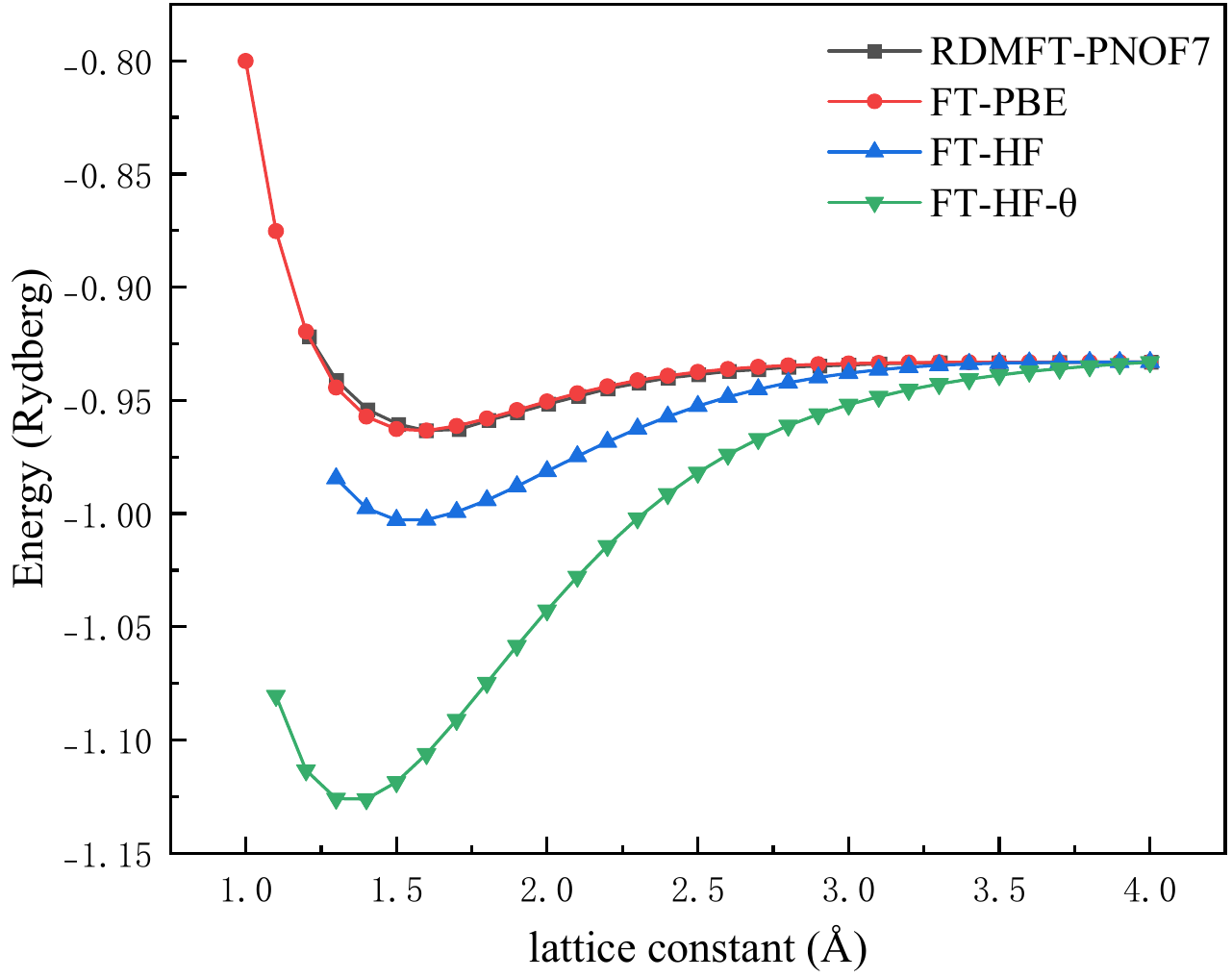}
\caption{
Per-atom dissociation energy of three-dimensional solid hydrogen as a function of the lattice constant, obtained from FT-PBE (and the exact-exchange-based FT-HF and FT-HF-$\theta$) in ABACUS, compared with the RDMFT-PNOF7 reference of Ref.~\cite{lew2025advancing} for a $10\times10\times10$ cluster of $1000$ hydrogen atoms. The temperature parameters used for FT-PBE, FT-HF, and FT-HF-$\theta$ are $\theta=0.122$, $0.16$, and $0.30$~Ry, respectively.
}
\label{fig:hsolid_ftpbe_rdmft}
\end{figure}

\subsection{Bulk modulus of V$_2$O$_3$}

Corundum-type V$_2$O$_3$ is a canonical strongly correlated transition-metal oxide and a benchmark system for studying Mott metal--insulator transition physics~\cite{imada1998metal,held2001mott,limelette2003universality}. Its electronic state is sensitive to pressure, temperature, and chemical substitution, and changes in the electronic structure are closely connected to the structural and elastic response. This makes the equation of state of V$_2$O$_3$ a useful test of correlation-sensitive bulk properties beyond molecular dissociation.

The sensitivity of the electronic structure is also reflected in the anomalous compression reported for corundum-type V$_2$O$_3$. Earlier experimental values of the bulk modulus range from approximately $172$ to $255$~GPa~\cite{mcwhan1970metal,sato1979hydrostatic,finger1980crystal,zhang2012pressure,nichols1981effect}, whereas a high-pressure x-ray-diffraction study reported $V_0^{\mathrm{exp}}=99.53$~\AA$^3$ for a cell containing two V$_2$O$_3$ formula units and $B_0=150.6$~GPa from a third-order Birch--Murnaghan fit with $B_0^{\prime}=4$ held fixed~\cite{ovsyannikov2013anomalous}. Because the available pressure--volume data did not allow an independent determination of $B_0^{\prime}$, the experimental $B_0^{\prime}=4$ is a fitting constraint rather than a measured value, and the reported $V_0$ and $B_0$ should be understood as conditional on this choice. In what follows we therefore treat the experimental analysis as providing a conditional reference pair, $B_0^{\mathrm{exp}}=150.6$~GPa with $B_0^{\prime}=4$.

The calculated energy--volume data were fitted to a four-parameter Murnaghan equation of state, from which the equilibrium volume $V_0$, bulk modulus $B_0$, and pressure derivative $B_0^{\prime}$ were obtained. Since these quantities are determined by the shape and curvature of the energy--volume curve, Fig.~\ref{fig:v2o3_ftldat_ev} first shows representative energy--volume curves at the characteristic temperature parameters used for FT-LDA and FT-LDA-$\theta$, together with the zero-temperature DFT-LDA curve. The FT-LDA, FT-LDA-$\theta$, and DFT-LDA curves give equilibrium volumes of $106.02$, $107.15$, and $91.24$~\AA$^3$, respectively: the two finite-temperature schemes give volumes slightly larger than the experimental $V_0^{\mathrm{exp}}=99.53$~\AA$^3$, whereas DFT-LDA gives a smaller one.

The dependence of the derived bulk properties on the temperature parameter is shown in Figs.~\ref{fig:v2o3_ftldat_b0} and \ref{fig:v2o3_ftldat_b0p}. At $\theta=0.107$~Ry, FT-LDA yields $B_0=151.29$~GPa and $B_0^{\prime}=5.02$; at $\theta=0.14$~Ry, FT-LDA-$\theta$ yields $B_0=151.49$~GPa and $B_0^{\prime}=4.83$; the zero-temperature DFT-LDA values are $B_0=275.87$~GPa and $B_0^{\prime}=4.33$. For comparison, Leonov \textit{et al.} reported a bulk modulus of $B_0\sim252$~GPa for V$_2$O$_3$ within non-magnetic GGA calculations~\cite{leonov2015metal}, reasonably close to our DFT-LDA result of $275.87$~GPa, indicating that conventional semilocal DFT likewise predicts a relatively stiff equation of state for V$_2$O$_3$. The DFT-LDA results here correspond to the non-magnetic phase, for which the standard local approximation misses the strong correlation among the V $3d$ electrons and describes an overly delocalized, over-bound state; this is the origin of the small equilibrium volume and the strongly overestimated bulk modulus. The finite-temperature occupation treatment substantially reduces this overestimate, bringing $B_0$ from $275.87$~GPa close to the experimental value. With the temperature parameter of each scheme chosen so that its $B_0$ approaches $150.6$~GPa, FT-LDA-$\theta$ gives $B_0^{\prime}=4.83$, closer to the adopted value of $4$ than the FT-LDA result of $5.02$; the inclusion of $E_{\theta}$ therefore leaves $B_0$ almost unchanged while improving the consistency of $B_0^{\prime}$ with the conditional reference pair. We note that the zero-temperature DFT-LDA value $B_0^{\prime}=4.33$ is numerically close to $4$ as well, although DFT-LDA severely overestimates $B_0$ itself. We emphasize that, within the present framework, this reduction of $B_0$ is of correlation origin rather than a thermal effect: the fractional occupations act as a description of the $3d$-electron correlation, softening the excessively strong bonding of the non-magnetic LDA description. This flattens the energy--volume curve on the large-volume side of the minimum and thereby lowers its curvature at equilibrium, i.e.\ the bulk modulus. The improvement in $B_0$ is thus of the same correlation origin as the improved dissociation behavior discussed in Secs.~\ref{sec:fthf_ftpbe0}--\ref{sec:hsolid}, not a temperature-induced (e.g.\ thermal-expansion) effect. Because the temperature parameters are selected separately for the different functionals, they need not coincide. Results obtained with the PBE and HSE functionals are presented in the Supplemental Material and exhibit behavior similar to that found with the functionals considered here.

\begin{figure}[t]
\centering
\includegraphics[width=0.95\columnwidth]{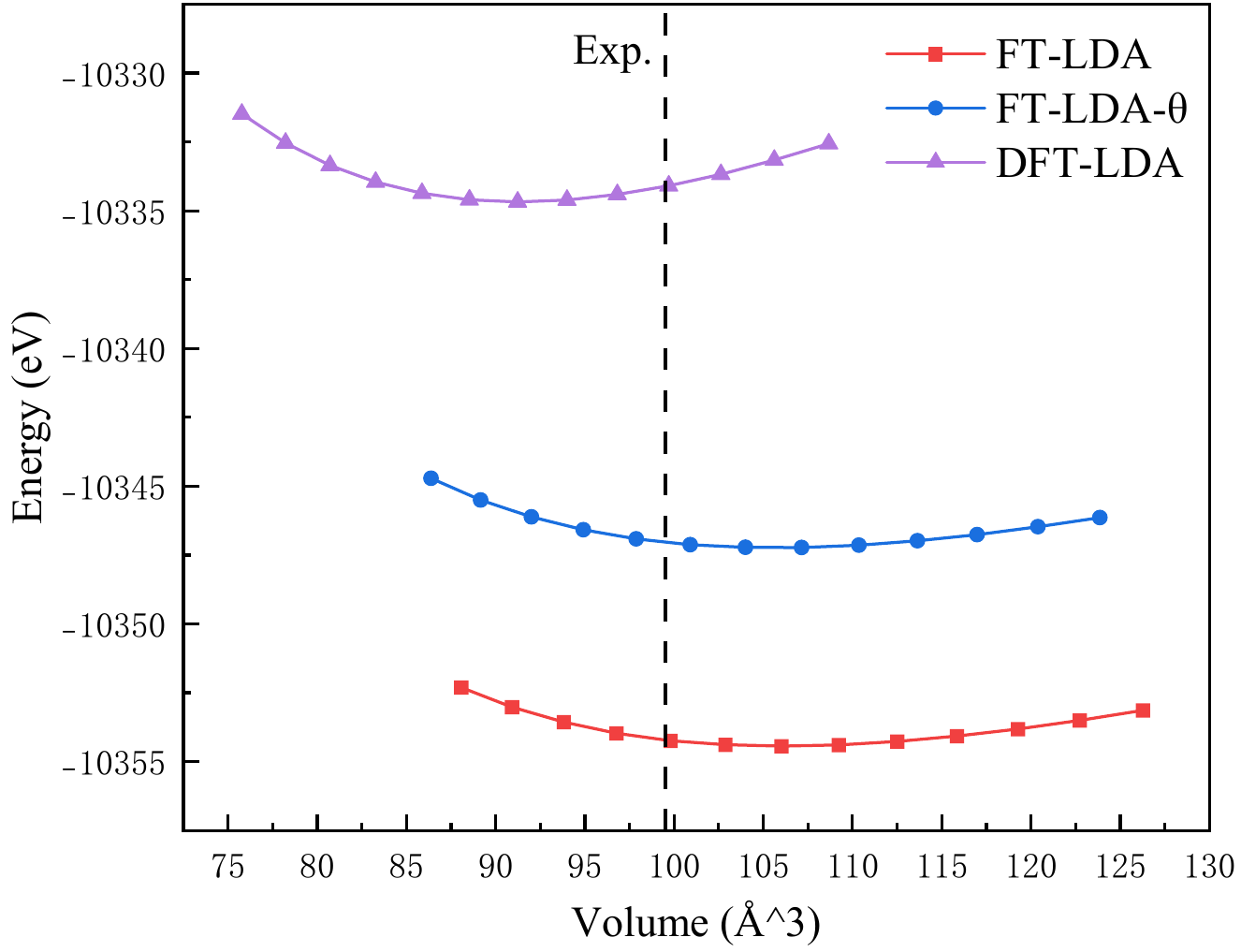}
\caption{
Energy--volume curves of corundum-type V$_2$O$_3$ obtained from FT-LDA at $\theta=0.107$~Ry, FT-LDA-$\theta$ at $\theta=0.14$~Ry, and zero-temperature DFT-LDA. The corresponding equilibrium volumes are $V_0=106.02$, $107.15$, and $91.24$~\AA$^3$, respectively. The vertical dashed line labeled ``Exp.'' marks the experimental zero-pressure equilibrium volume $V_0^{\mathrm{exp}}=99.53$~\AA$^3$~\cite{ovsyannikov2013anomalous}.
}
\label{fig:v2o3_ftldat_ev}
\end{figure}

\begin{figure*}[tbp]
\centering
\subfloat[$B_0$]{%
    \includegraphics[width=0.42\textwidth]{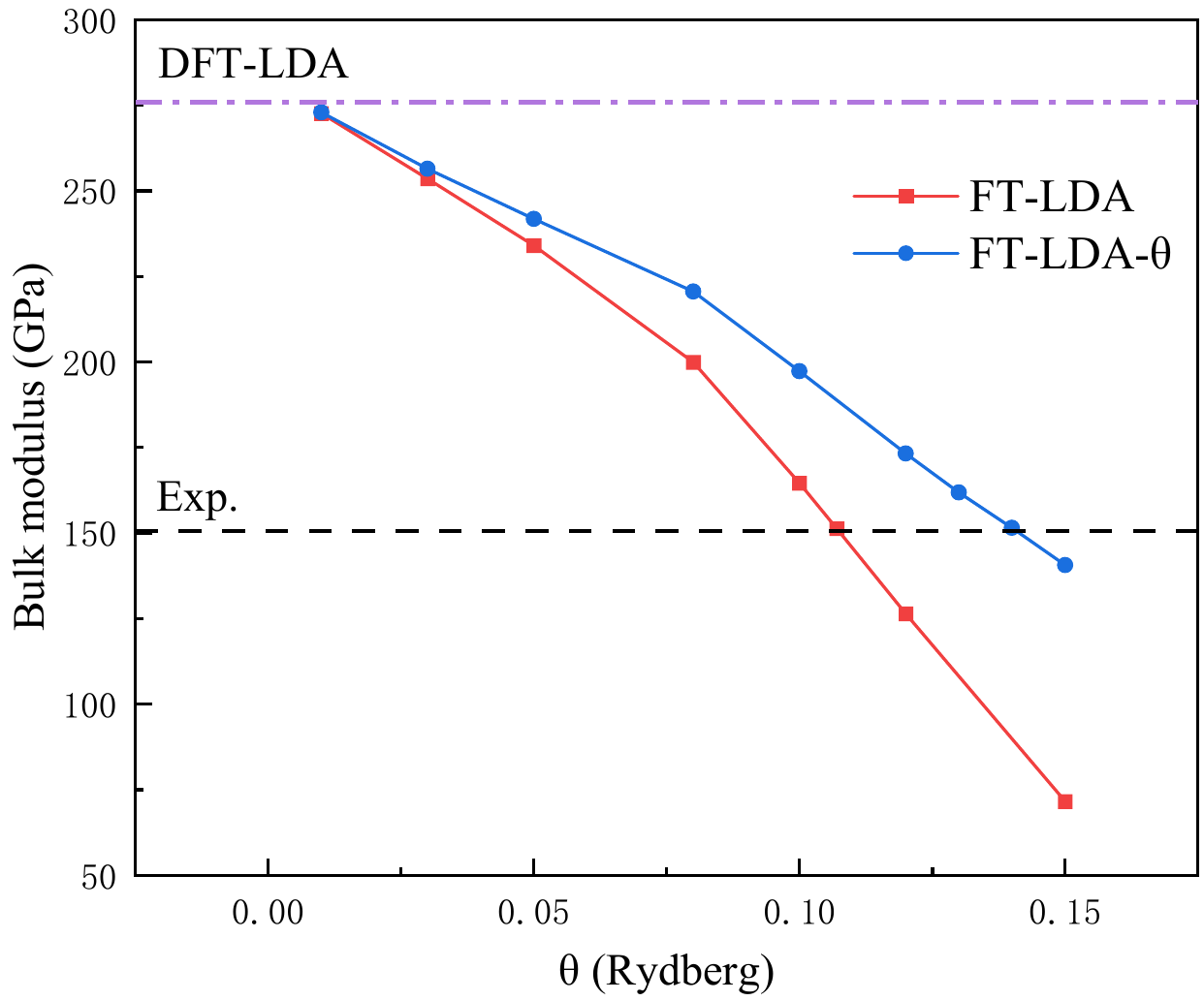}
    \label{fig:v2o3_ftldat_b0}
}
\hspace{0.03\textwidth}
\subfloat[$B_0^{\prime}$]{%
    \includegraphics[width=0.42\textwidth]{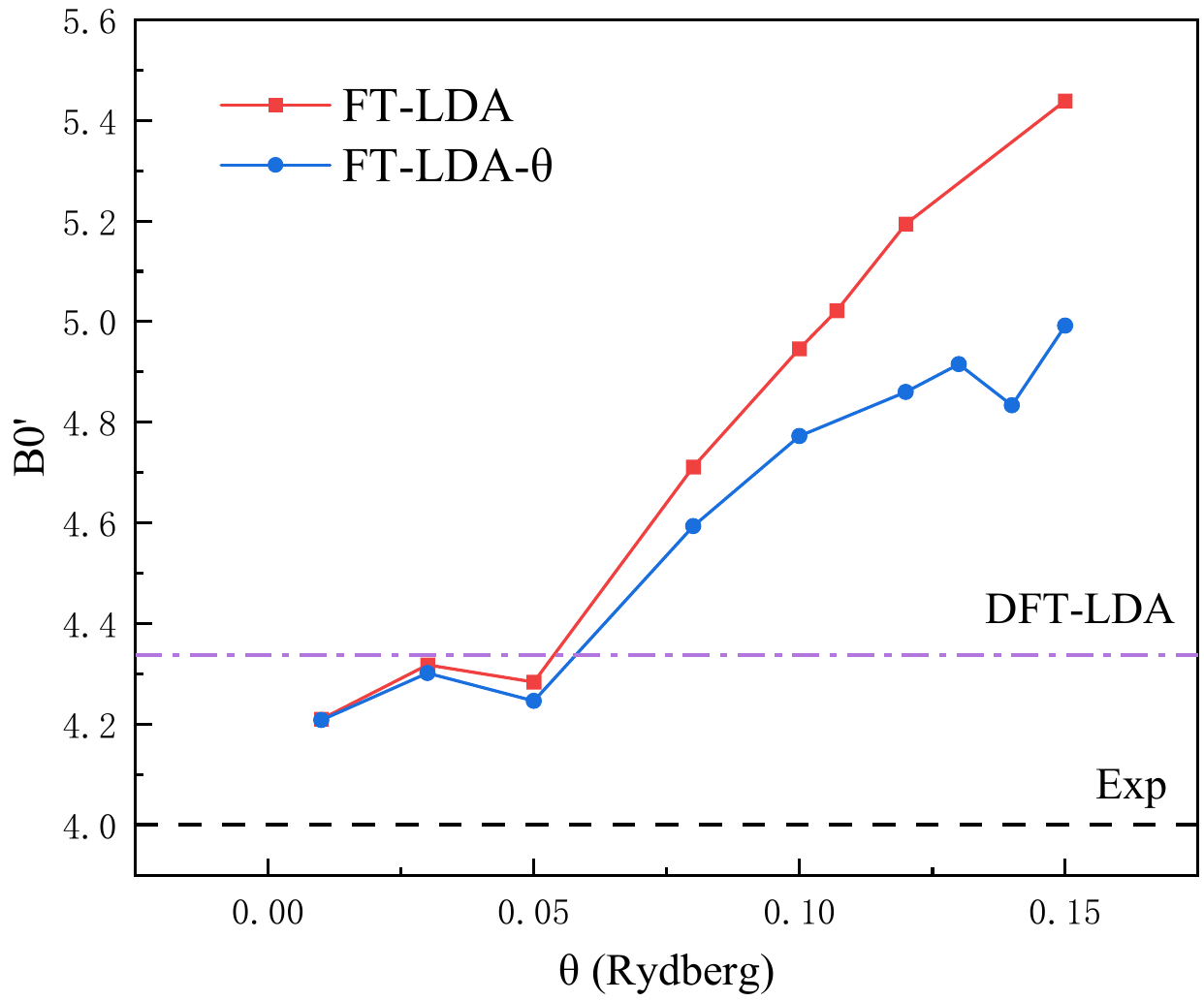}
    \label{fig:v2o3_ftldat_b0p}
}
\caption{
Dependence of (a) the bulk modulus $B_0$ and (b) its pressure derivative $B_0^{\prime}$ on the temperature parameter $\theta$ for FT-LDA and FT-LDA-$\theta$ in corundum-type V$_2$O$_3$. The black dashed lines labeled ``Exp'' indicate $B_0=150.6$~GPa and $B_0^{\prime}=4$, with the latter fixed in the cited experimental equation-of-state fit~\cite{ovsyannikov2013anomalous}. The purple dashed lines labeled ``DFT-LDA'' indicate the corresponding zero-temperature DFT-LDA values, $B_0=275.87$~GPa and $B_0^{\prime}=4.33$. The FT-LDA values at $\theta=0.107$~Ry are $B_0=151.29$~GPa and $B_0^{\prime}=5.02$, while the FT-LDA-$\theta$ values at $\theta=0.14$~Ry are $B_0=151.49$~GPa and $B_0^{\prime}=4.83$.
}
\label{fig:v2o3_bulk_ftldat}
\end{figure*}

\section{Discussion}
\label{sec:discussion}

The present results clarify the role of FT-AC-RDMFT at both the formal and practical levels. Formally, FT-AC-RDMFT provides a natural variational framework in which i-DMFT and TAO-DFT can be understood as two reduced forms of a common underlying theory. Within this framework, the effective equation for the natural orbitals and the Fermi--Dirac form of the occupation numbers follow directly from the stationarity conditions of the 1-RDM functional. The finite-temperature occupation structure is therefore not introduced empirically, but emerges naturally from the variational principle itself.

From a computational perspective, the present formulation shares the equation structure of established finite-temperature mean-field methods. Once a specific functional approximation is adopted, the working equations take the same form as those of Hartree--Fock or Kohn--Sham DFT, so that the natural orbitals and Fermi--Dirac occupations can be obtained with the same standard techniques and at comparable cost per step, with the computational cost determined mainly by the underlying exchange-correlation treatment. This favorable structure is shared by finite-temperature occupation-based methods such as i-DMFT~\cite{wang2022scf}, and is a property of the formulation rather than a specific achievement of the present work.

The numerical results further show that the beneficial role of finite-temperature occupations is more general than an exact-exchange-based interpretation alone would suggest. The equivalence between i-DMFT and FT-HF identifies FT-HF as the natural exact-exchange realization of the reduced FT-AC-RDMFT framework. At the same time, the comparison between FT-HF and FT-PBE0, together with the FT-PBE results for solid hydrogen, indicates that the improvement in dissociation behavior is not tied exclusively to exact exchange. Rather, the finite-temperature occupation structure itself appears to be the essential ingredient, while the quantitative performance remains functional dependent.

Beyond this common finite-temperature occupation structure, the FT-AC-RDMFT framework leaves open several possible routes to go beyond FT-HF. The $E_{\theta}$ term is one such route: it is not required by the FT-RDMFT variational framework or by the adiabatic connection itself, but is introduced specifically to make contact with TAO-DFT-type, density-based formulations. In this sense, exploring $E_{\theta}$ is a step along one particular direction rather than a necessary ingredient of the framework.

Our exploration of this route gives a mixed picture. On the one hand, the current LDA-based treatment of $E_{\theta}$, combined with LDA correlation, does not improve the dissociation curves of the molecules studied here, nor the dissociation of solid hydrogen; the present local form appears not to be well matched to the strongly inhomogeneous, orbital-dependent electronic structure of stretched systems. On the other hand, the V$_2$O$_3$ results indicate that the effect of $E_{\theta}$ is not uniformly negative: the finite-temperature occupations bring the strongly overestimated non-magnetic LDA bulk modulus close to experiment, and while the inclusion of $E_{\theta}$ leaves $B_0$ itself nearly unchanged, the FT-LDA-$\theta$ equation-of-state parameters are somewhat more consistent with the conditional experimental reference pair. Importantly, within the present framework this improvement is of the same correlation origin as the improved dissociation behavior---a softening of the overly strong LDA bonding by the fractional occupations---rather than a thermal effect. The quantitative impact of $E_{\theta}$ thus depends on both the property under consideration and the quality of the approximation used, and improving its form for orbital-dependent and inhomogeneous systems is one concrete direction for future work---one that concerns this particular route rather than the FT-RDMFT framework itself.

A further open issue concerns the temperature parameter $\theta$ itself. In the present framework $\theta$ is treated as an adjustable parameter, chosen empirically for each system and functional, and a first-principles prescription for determining it is still lacking. While the adiabatic-connection analysis clarifies the physical meaning of the entropy coefficient by relating it to the characteristic temperature $\theta_c$ that reproduces the interacting 1-RDM (with $\kappa=\theta_c/2$ at first order), determining $\theta_c$ from first principles remains an open problem. Establishing such a prescription would place the connection to the i-DMFT coefficient $\kappa$ on a fully first-principles footing, and is an important direction for future work.

\section{Summary}
\label{sec:summary}

In summary, by combining the 1-RDM variational framework of RDMFT with a finite-temperature non-interacting reference, we have developed the FT-AC-RDMFT framework, which places i-DMFT and TAO-DFT in a common 1-RDM-based setting: i-DMFT corresponds to the exact-exchange-based reduction, while TAO-DFT emerges when exact exchange is replaced by a density-based exchange-correlation functional. The effective eigenvalue equation and the Fermi--Dirac occupations follow directly from the variational principle, and the self-consistent calculations show that finite-temperature occupations can improve dissociation behavior in both exact-exchange-based and density-functional-based schemes. The present work thus provides both a unified theoretical perspective and a practical starting point for future developments of finite-temperature reduced density-matrix functional theory.

\section*{Data Availability}

The numerical data for all calculations reported in this work are publicly available at the data repository~\cite{ftacrdmft_data}. The source code of the ABACUS software used in this work is publicly available in the GitHub repository~\cite{abacus_e_theta_lda}.

\begin{acknowledgments}
 We would like to thank Prof. Zhong Fang and Prof. Jian Wang for their helpful discussions. We acknowledge the funding support from the National Natural Science Foundation of China (Grant Nos. 12188101, 12374067, 12134012) and the Advanced Materials-National Science and Technology Major Project (Grant No. 2025ZD0618401). H. W. acknowledges support from the New Cornerstone Science Foundation through the XPLORER PRIZE. This work was also supported by the Strategic Priority Research Program of Chinese Academy of Sciences under Grant No. XDB0500201, the National Key Research and Development Program of China (Grant Nos. 2022YFA1403800 and 2023YFA1507004), and the robotic AI-Scientist platform of the Chinese Academy of Sciences. 
\end{acknowledgments}

\endgroup

\bibliography{reference}

\end{document}


\raggedbottom

\title{Supplemental materials for \\ Finite-temperature adiabatic-connection reduced density-matrix functional theory: a unified framework for i-DMFT and TAO-DFT
}

\author{Jingang Han}
\affiliation{Institute of Physics, Chinese Academy of Sciences, Beijing 100190, China}
\affiliation{School of Physical Sciences, University of Chinese Academy of Sciences, Beijing 100049, China}
\author{Huanjing Gong}
\affiliation{Institute of Physics, Chinese Academy of Sciences, Beijing 100190, China}
\affiliation{School of Physical Sciences, University of Chinese Academy of Sciences, Beijing 100049, China}
\author{Kai Luo}
\email{kluo@njust.edu.cn}
\affiliation{Department of Applied Physics, Nanjing University of Science and Technology, Nanjing 210094, China}
\author{Hongming Weng}
\email{hmweng@iphy.ac.cn}
\affiliation{Institute of Physics, Chinese Academy of Sciences, Beijing 100190, China}
\author{Xinguo Ren}
\email{renxg@iphy.ac.cn}
\affiliation{Institute of Physics, Chinese Academy of Sciences, Beijing 100190, China}

\newcommand{\XR}[1]{\color{red}{\bf #1 }}
\newcommand{\XB}[1]{\color{blue}{ #1 }}
\newcommand{\XY}[1]{\color{yellow}{\bf #1 }}

\maketitle

\section{Additional results for the dissociation of solid hydrogen}

In this section, we provide additional results for the dissociation of solid hydrogen as a function of the lattice constant, using the DoNOF-PNOF7 data of Ref.~\cite{lew2025advancing} as the reference. Besides the FT-PBE result discussed in the main text, we examine FT-HF, FT-HSE, and FT-HF-$\theta$.

\subsection{FT-HF and FT-HSE}

\begin{figure}[!htbp]
\centering
\subfloat[FT-HF]{%
    \includegraphics[width=0.90\columnwidth]{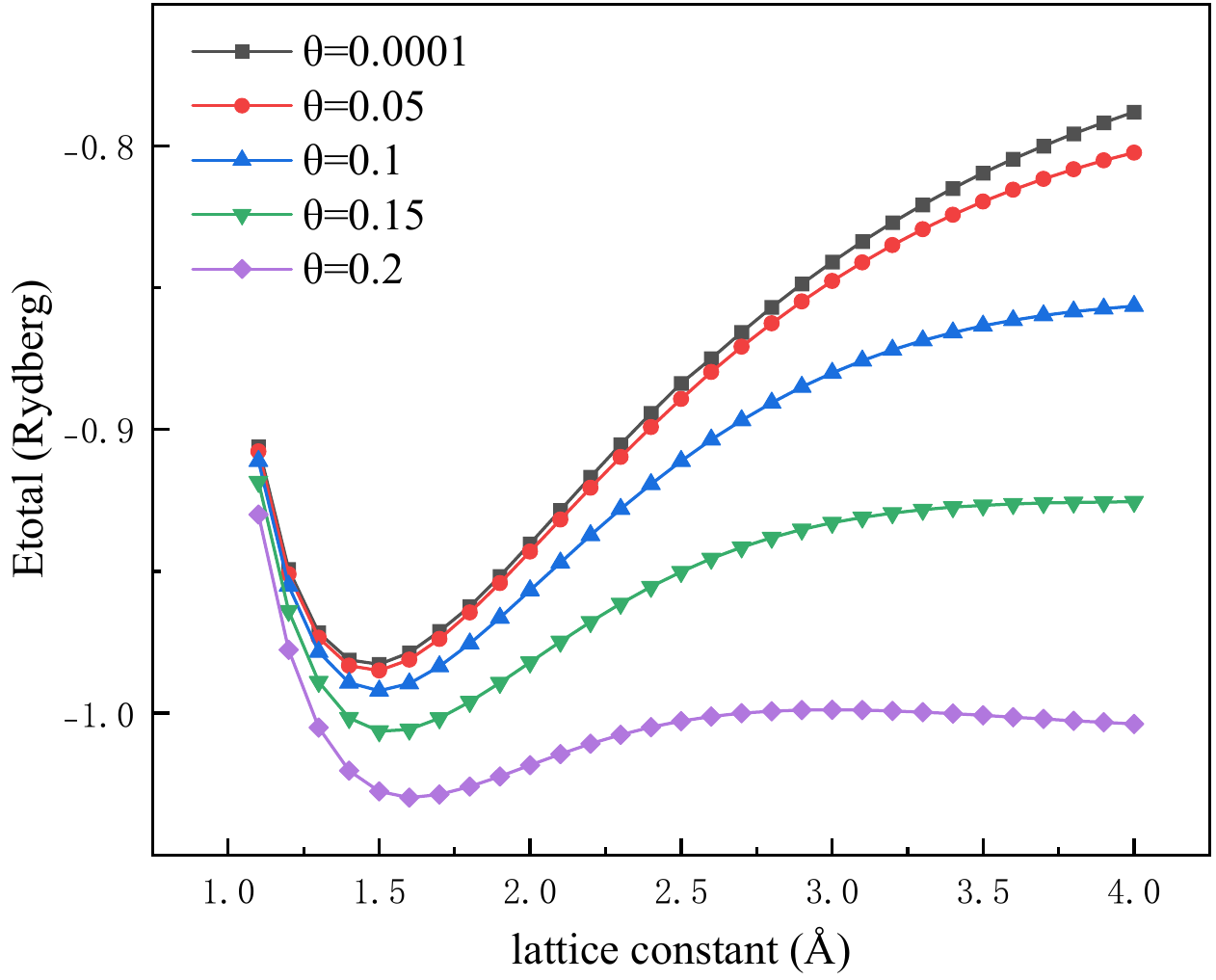}
    \label{fig:sm_hsolid_fthf_scan}
}
\par\vspace{0.8em}
\subfloat[FT-HSE]{%
    \includegraphics[width=0.90\columnwidth]{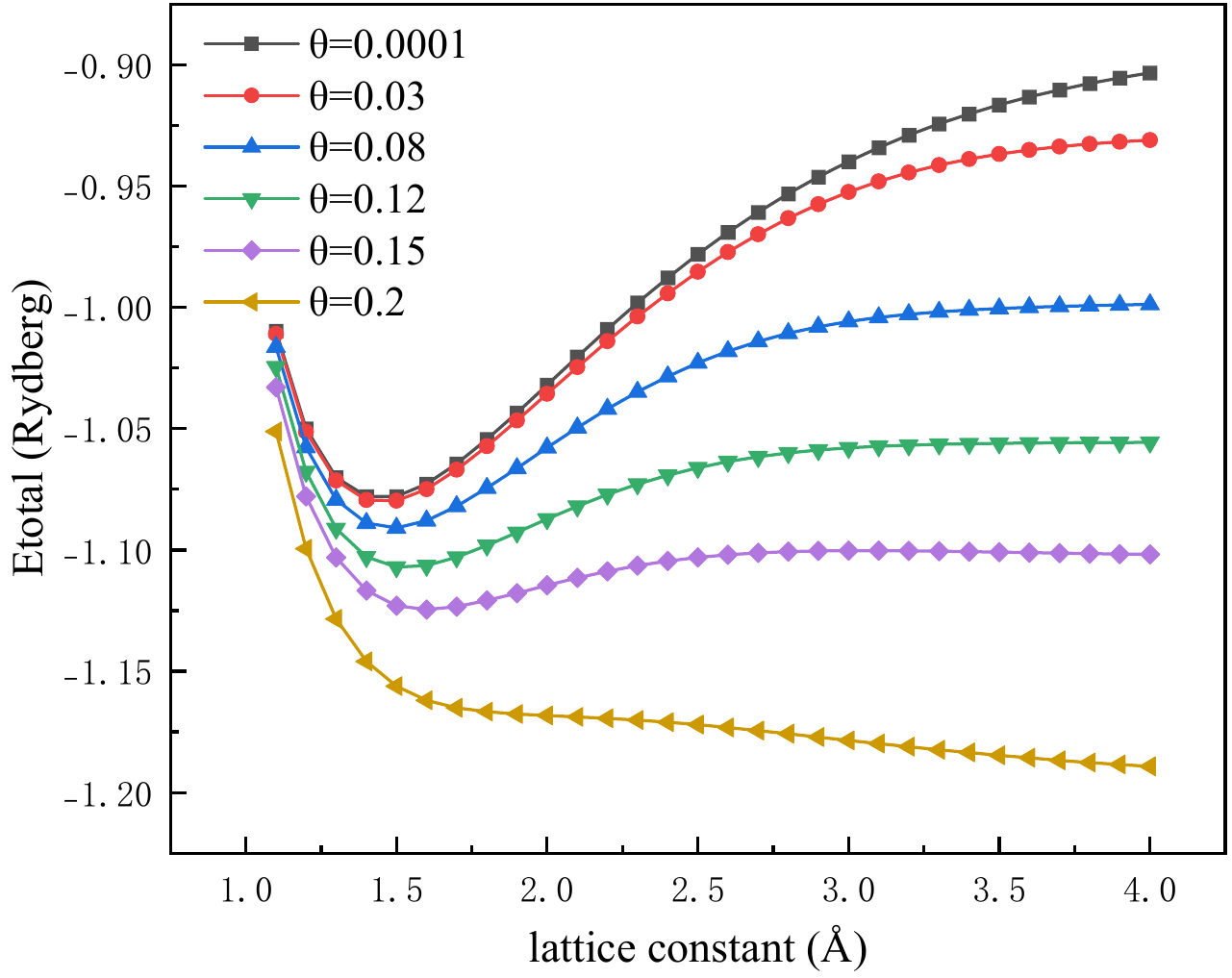}
    \label{fig:sm_hsolid_fthse_scan}
}
\caption{
Representative dissociation curves of solid hydrogen obtained from (a) FT-HF and (b) FT-HSE at different temperature parameters.
}
\label{fig:sm_hsolid_fthf_fthse_scan}
\end{figure}

Representative FT-HF and FT-HSE dissociation curves for solid hydrogen are shown in Fig.~\ref{fig:sm_hsolid_fthf_fthse_scan}, where the left panel corresponds to FT-HF and the right panel to FT-HSE. For FT-HF, a temperature around $\theta \approx 0.16$~Ry gives the most reasonable dissociation profile. For FT-HSE, the corresponding value is about $\theta \approx 0.13$~Ry. In both cases, if the temperature is increased further, the dissociation curve develops a downward trend in the large-distance region. Nevertheless, within a suitable temperature range, both FT-HF and FT-HSE still recover the correct qualitative physical picture of dissociation.

\begin{figure}[!htbp]
\centering
\includegraphics[width=0.90\columnwidth]{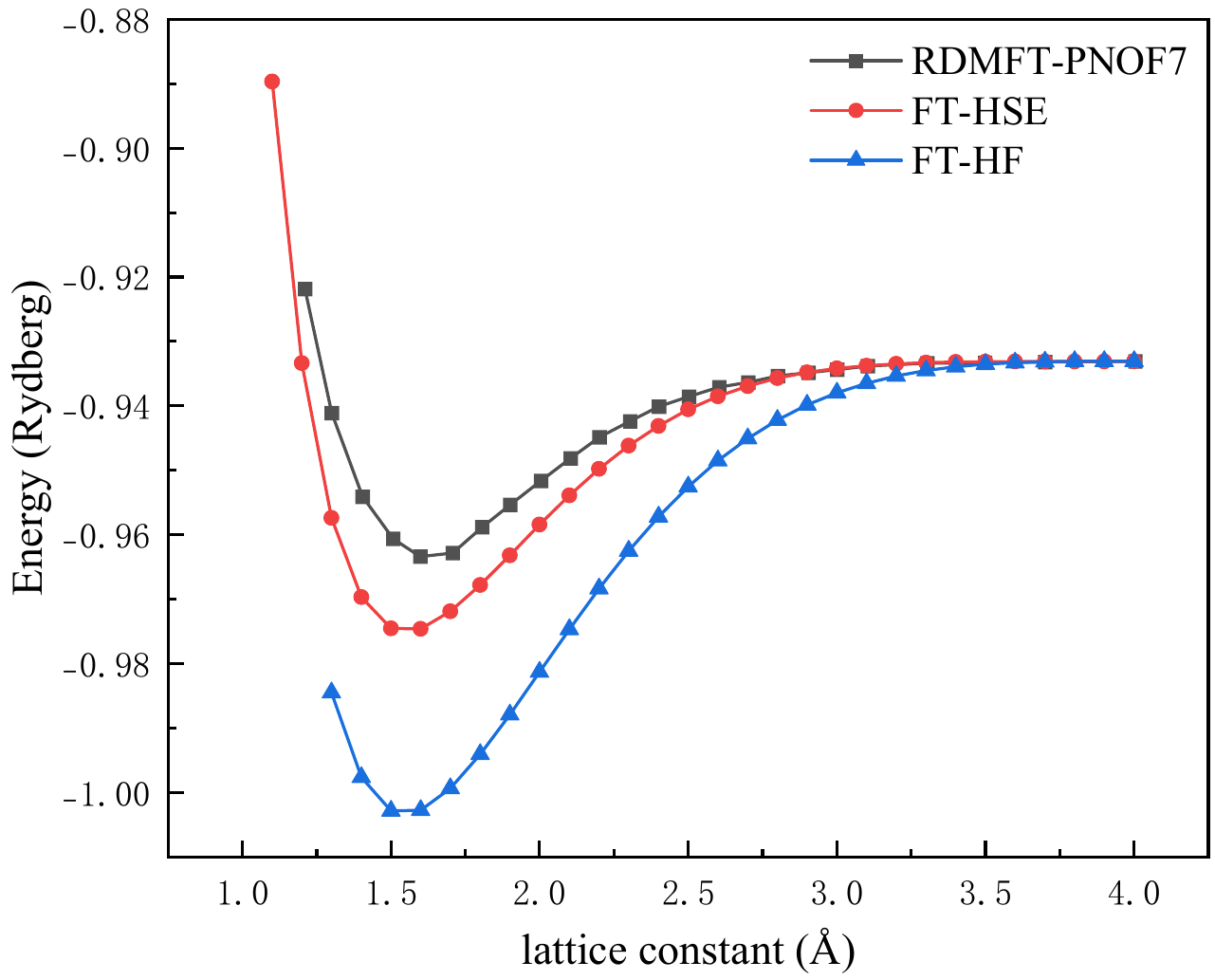}
\caption{
Comparison of the dissociation curves of solid hydrogen obtained from FT-HSE ($\theta=0.13$~Ry), FT-HF ($\theta=0.16$~Ry), and the DoNOF-PNOF7 reference from Ref.~\cite{lew2025advancing}. The reference curve is shifted by a constant.
}
\label{fig:sm_hsolid_compare_fthf_fthse_rdmft}
\end{figure}

Figure~\ref{fig:sm_hsolid_compare_fthf_fthse_rdmft} compares the FT-HF and FT-HSE curves at these representative temperatures with the DoNOF-PNOF7 reference. Although both methods reproduce the overall dissociation behavior, their dissociation energies remain higher than the reference value. Therefore, within the present parameter range, neither FT-HF nor FT-HSE reproduces the DoNOF-PNOF7 dissociation curve quantitatively.

\subsection{FT-HF-$\theta$}

\begin{figure}[!htbp]
\centering
\subfloat[]{%
    \includegraphics[width=0.90\columnwidth]{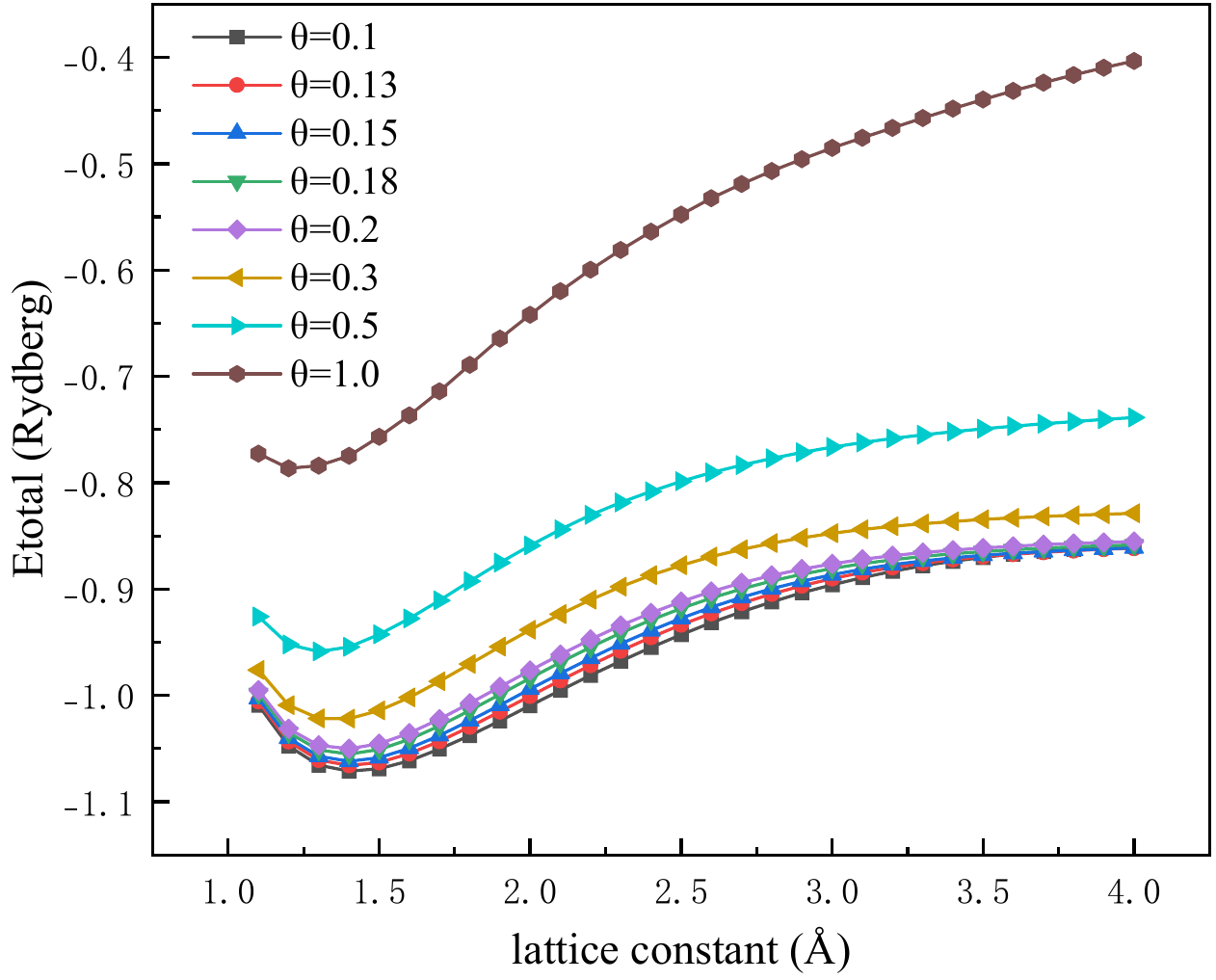}
    \label{fig:sm_hsolid_fthf_theta_etotal}
}
\par\vspace{0.8em}
\subfloat[]{%
    \includegraphics[width=0.90\columnwidth]{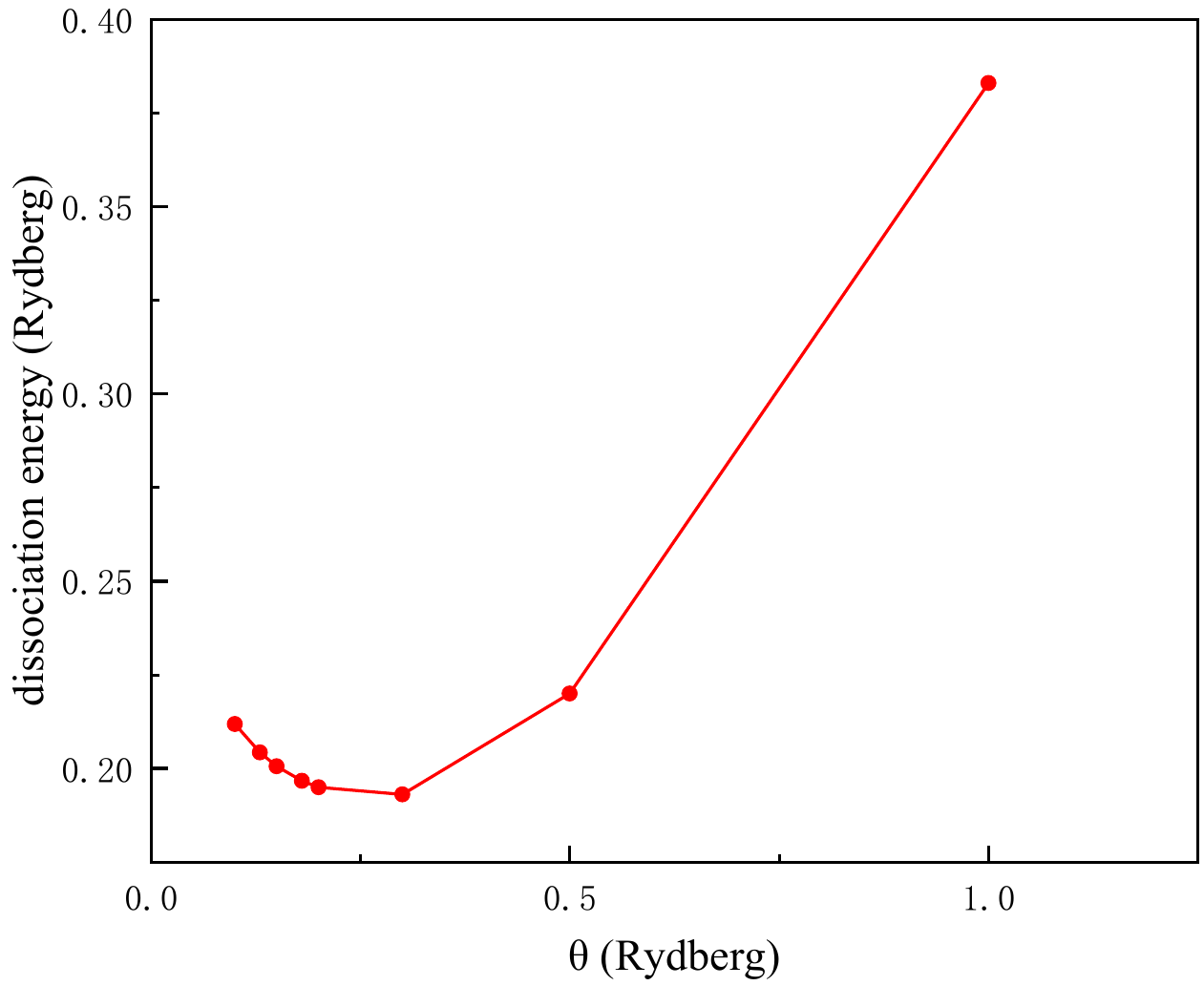}
    \label{fig:sm_hsolid_fthf_theta_de}
}
\caption{
Dissociation curves and dissociation energies of solid hydrogen obtained from FT-HF-$\theta$ at different temperatures.
}
\label{fig:sm_hsolid_fthf_theta}
\end{figure}

We also tested FT-HF-$\theta$, i.e., the FT-AC-RDMFT scheme with exact exchange, LDA correlation, and the present $E_{\theta}$ approximation. As shown in Fig.~\ref{fig:sm_hsolid_fthf_theta}, the dissociation energy varies only weakly with temperature and remains close to $0.2$~Ry or higher over the range considered, well above the reference value. Thus, unlike the molecular cases, the present FT-HF-$\theta$ implementation does not improve the dissociation of solid hydrogen. As in the main text, this points to the current approximations to $E_c^{\mathrm{LDA}}$ and $E_{\theta}$, rather than to the FT-AC-RDMFT formalism itself, as the source of the difficulty.

\FloatBarrier
\section{Additional results for the bulk modulus of V$_2$O$_3$}

In this section, we provide additional finite-temperature results for the bulk modulus of corundum-type V$_2$O$_3$. Owing to the higher cost of the hybrid-functional calculations, uniform Monkhorst--Pack meshes of $10\times10\times10$ and $6\times6\times6$ were used for FT-PBE and FT-HSE, respectively. As a reference, the experimental equation-of-state fit gives $B_0=150.6$~GPa with $B_0^{\prime}=4$ held fixed~\cite{ovsyannikov2013anomalous}. For each temperature, an energy--volume curve is generated around the relaxed structure, and $B_0$ and $B_0^{\prime}$ are extracted from the fit.

\subsection{FT-PBE}

\begin{figure}[H]
\centering
\includegraphics[width=0.90\columnwidth]{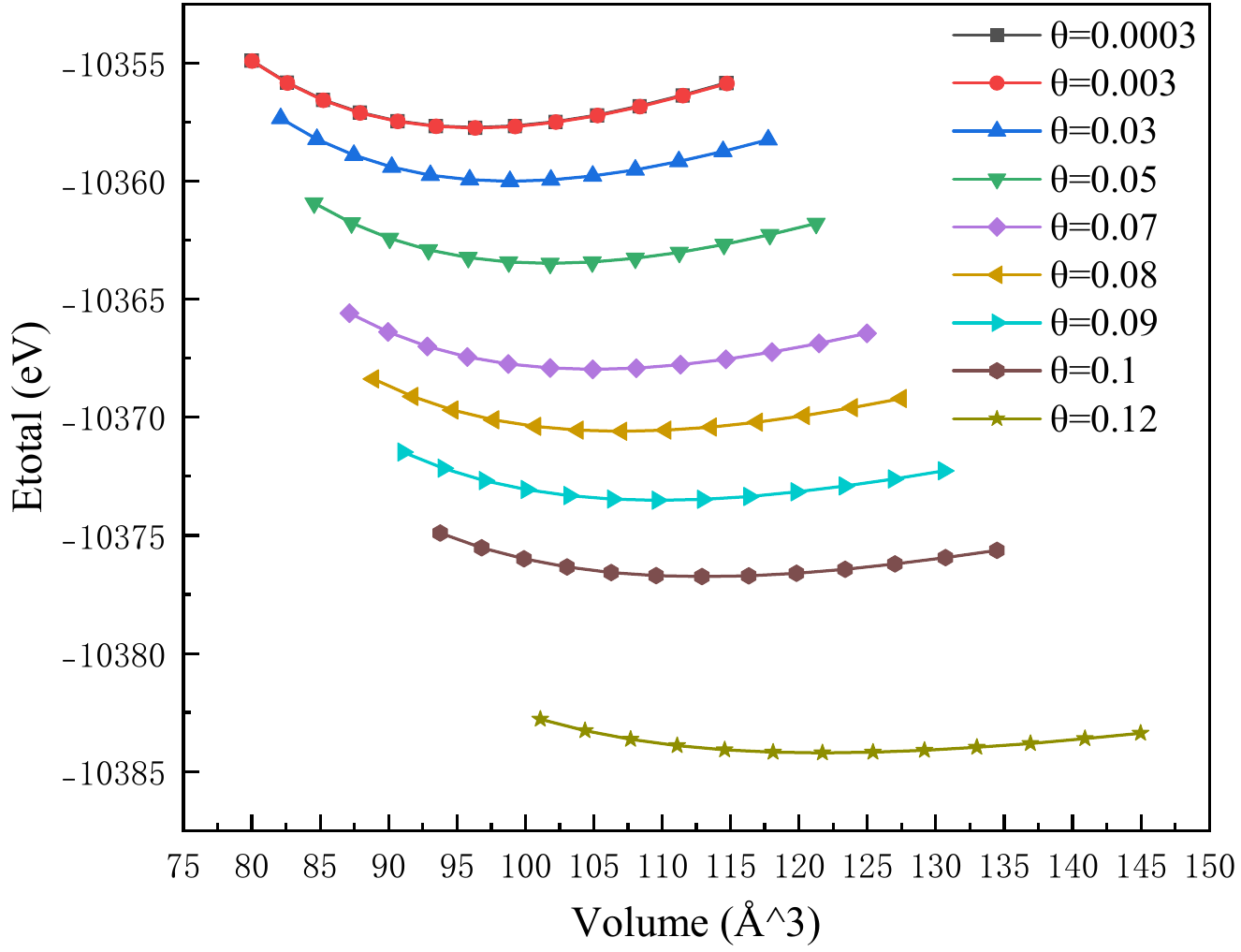}
\caption{
Energy--volume curves of corundum-type V$_2$O$_3$ obtained from FT-PBE at different temperatures.
}
\label{fig:sm_v2o3_ftpbe_ev}
\end{figure}

\begin{figure}[H]
\centering
\subfloat[$B_0$]{
    \includegraphics[width=0.90\columnwidth]{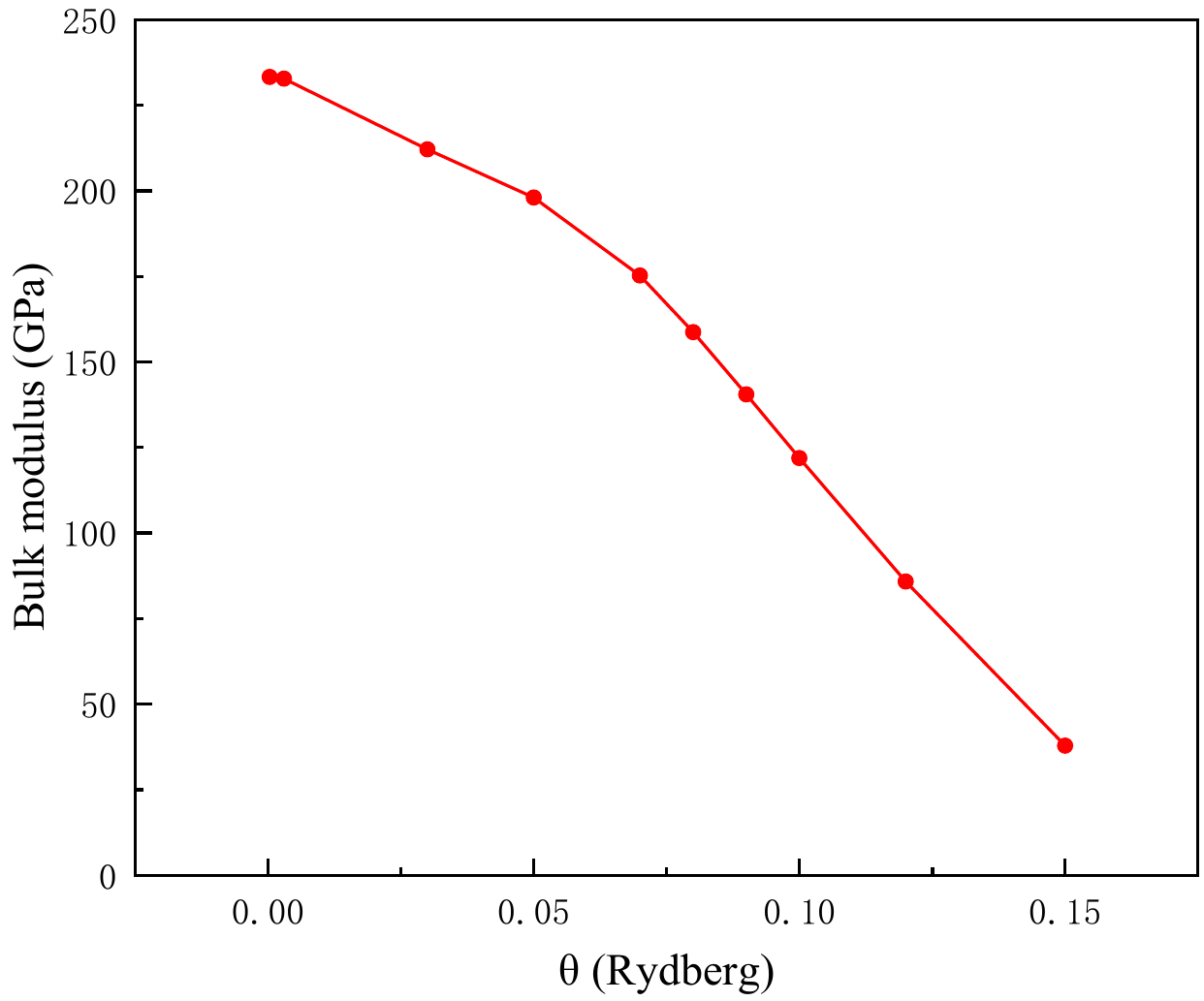}
    \label{fig:sm_v2o3_ftpbe_b0_a}
}
\par\vspace{0.8em}
\subfloat[$B_0^{\prime}$]{
    \includegraphics[width=0.90\columnwidth]{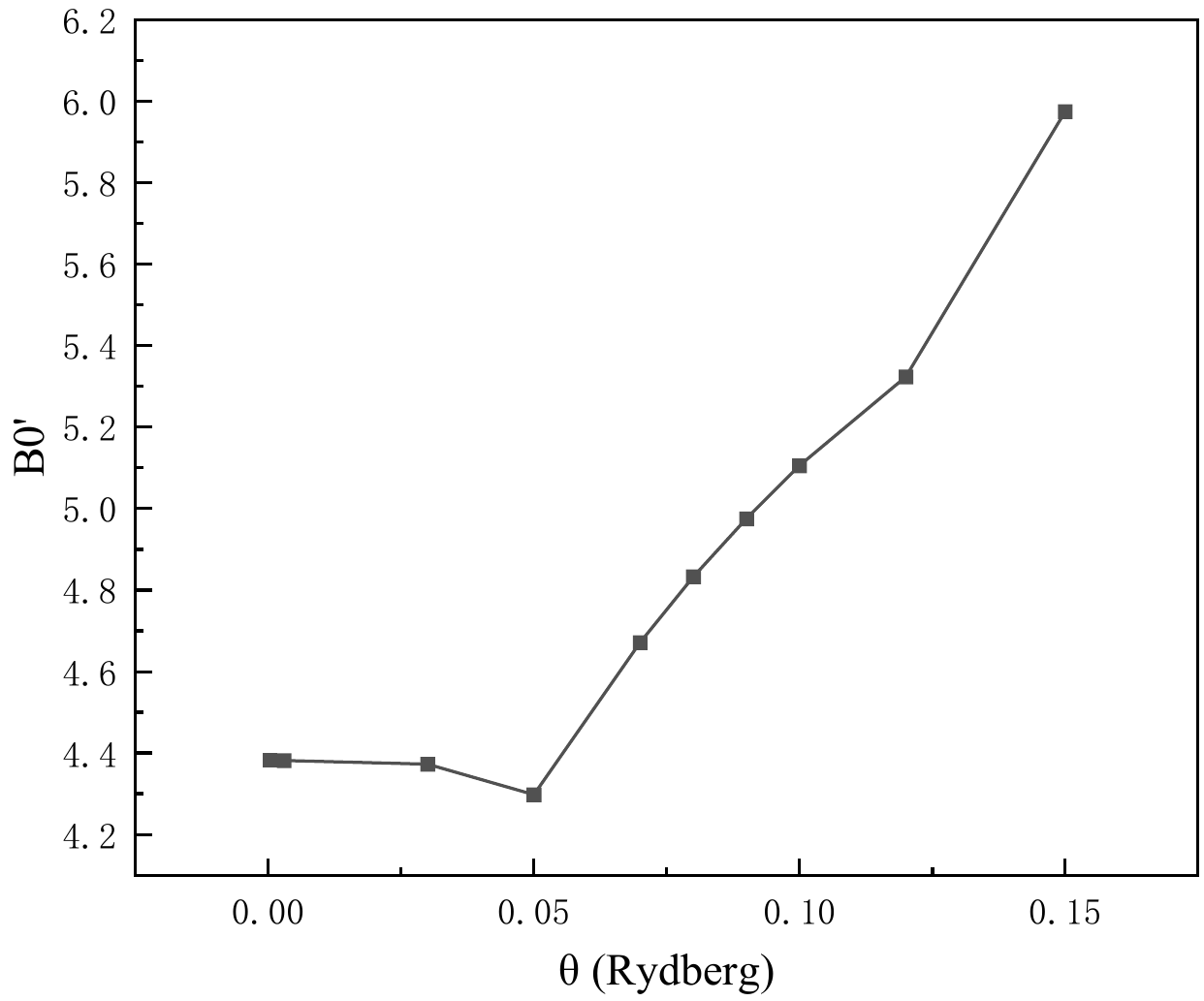}
    \label{fig:sm_v2o3_ftpbe_b0_b}
}
\caption{
Bulk modulus $B_0$ and pressure derivative $B_0^{\prime}$ of corundum-type V$_2$O$_3$ obtained from FT-PBE at different temperatures. At $\theta=0.08$~Ry, FT-PBE gives $B_0=158.69$~GPa and $B_0^{\prime}=4.83$.
}
\label{fig:sm_v2o3_ftpbe_b0}
\end{figure}

Figure~\ref{fig:sm_v2o3_ftpbe_ev} shows the energy--volume curves obtained from FT-PBE at different temperatures. From these curves, the corresponding bulk modulus $B_0$ and pressure derivative $B_0^{\prime}$ are estimated, as shown in Fig.~\ref{fig:sm_v2o3_ftpbe_b0}. At $\theta=0.08$~Ry, FT-PBE gives $B_0=158.69$~GPa and $B_0^{\prime}=4.83$. Thus, FT-PBE yields a reasonable pressure derivative, but the bulk modulus remains somewhat larger than the experimental value.

\vspace{1.0em}

\subsection{FT-HSE}

\begin{figure}[H]
\centering
\subfloat[$B_0$]{%
    \includegraphics[width=0.90\columnwidth]{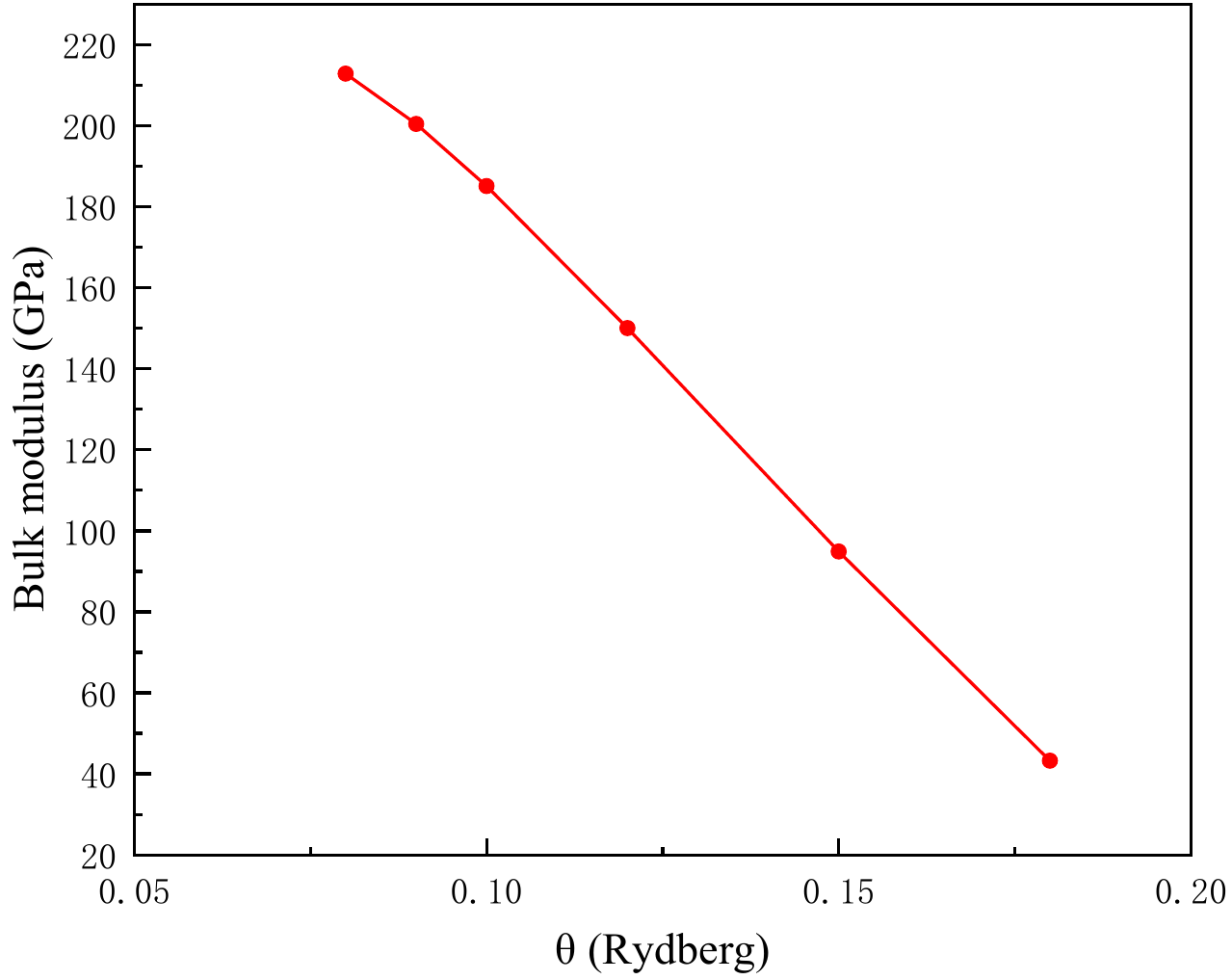}
    \label{fig:sm_v2o3_fthse_b0_a}
}
\par\vspace{0.8em}
\subfloat[$B_0^{\prime}$]{%
    \includegraphics[width=0.90\columnwidth]{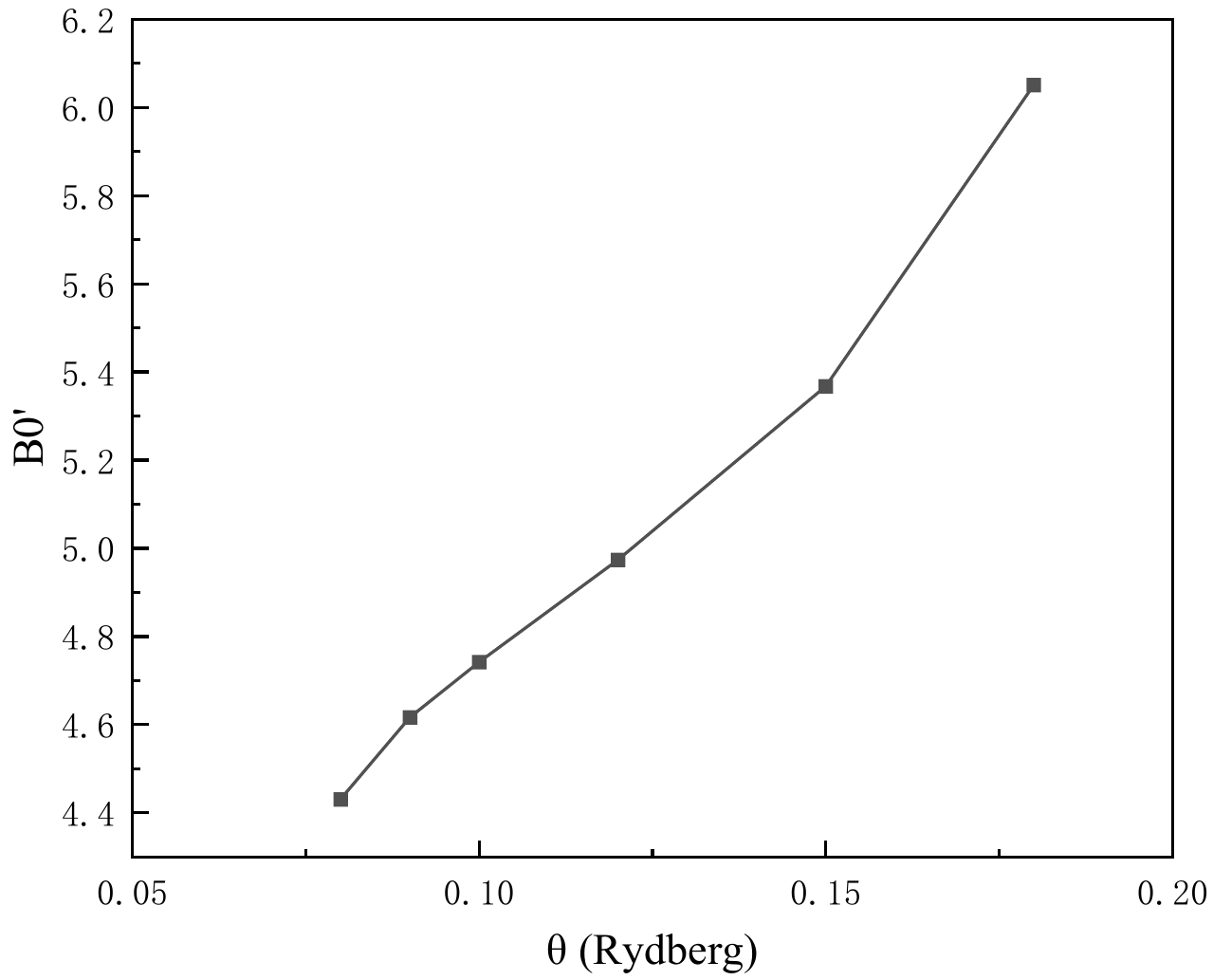}
    \label{fig:sm_v2o3_fthse_b0_b}
}
\caption{
Bulk modulus $B_0$ and pressure derivative $B_0^{\prime}$ of corundum-type V$_2$O$_3$ obtained from FT-HSE at different temperatures. At $\theta=0.12$~Ry, FT-HSE gives $B_0=150.02$~GPa and $B_0^{\prime}=4.97$.
}
\label{fig:sm_v2o3_fthse_b0}
\end{figure}

The corresponding FT-HSE results are summarized in Fig.~\ref{fig:sm_v2o3_fthse_b0}. At $\theta=0.12$~Ry, FT-HSE gives $B_0=150.02$~GPa and $B_0^{\prime}=4.97$. In this case, the bulk modulus is in very good agreement with experiment, while the pressure derivative remains somewhat larger than the value fixed in the experimental EOS fit.

\FloatBarrier
\bibliography{reference}